\documentclass[twocolumn]{aastex701}

\usepackage{amsmath}
\usepackage{multirow} 

\makeatletter
\let\jnl@style=\rm
\def\ref@jnl#1{{\jnl@style#1}}

\def\aj{\ref@jnl{AJ}}                   
\def\actaa{\ref@jnl{Acta Astron.}}      
\def\araa{\ref@jnl{ARA\&A}}             
\def\apj{\ref@jnl{ApJ}}                 
\def\apjl{\ref@jnl{ApJ}}                
\def\apjs{\ref@jnl{ApJS}}               
\def\ao{\ref@jnl{Appl.~Opt.}}           
\def\apss{\ref@jnl{Ap\&SS}}             
\def\aap{\ref@jnl{A\&A}}                
\def\aapr{\ref@jnl{A\&A~Rev.}}          
\def\aaps{\ref@jnl{A\&AS}}              
\def\azh{\ref@jnl{AZh}}                 
\def\baas{\ref@jnl{BAAS}}               
\def\bac{\ref@jnl{Bull. astr. Inst. Czechosl.}}
\def\caa{\ref@jnl{Chinese Astron. Astrophys.}}
\def\cjaa{\ref@jnl{Chinese J. Astron. Astrophys.}}
\def\icarus{\ref@jnl{Icarus}}           
\def\jcap{\ref@jnl{J. Cosmology Astropart. Phys.}}
\def\jrasc{\ref@jnl{JRASC}}             
\def\memras{\ref@jnl{MmRAS}}            
\def\mnras{\ref@jnl{MNRAS}}             
\def\na{\ref@jnl{New A}}                
\def\nar{\ref@jnl{New A Rev.}}          
\def\pra{\ref@jnl{Phys.~Rev.~A}}        
\def\prb{\ref@jnl{Phys.~Rev.~B}}        
\def\prc{\ref@jnl{Phys.~Rev.~C}}        
\def\prd{\ref@jnl{Phys.~Rev.~D}}        
\def\pre{\ref@jnl{Phys.~Rev.~E}}        
\def\prl{\ref@jnl{Phys.~Rev.~Lett.}}    
\def\pasa{\ref@jnl{PASA}}               
\def\pasp{\ref@jnl{PASP}}               
\def\pasj{\ref@jnl{PASJ}}               
\def\rmxaa{\ref@jnl{Rev. Mexicana Astron. Astrofis.}}%
\def\qjras{\ref@jnl{QJRAS}}             
\def\skytel{\ref@jnl{S\&T}}             
\def\solphys{\ref@jnl{Sol.~Phys.}}      
\def\sovast{\ref@jnl{Soviet~Ast.}}      
\def\ssr{\ref@jnl{Space~Sci.~Rev.}}     
\def\zap{\ref@jnl{ZAp}}                 
\def\nat{\ref@jnl{Nature}}              
\def\iaucirc{\ref@jnl{IAU~Circ.}}       
\def\aplett{\ref@jnl{Astrophys.~Lett.}} 
\def\apspr{\ref@jnl{Astrophys.~Space~Phys.~Res.}}
\def\bain{\ref@jnl{Bull.~Astron.~Inst.~Netherlands}} 
\def\fcp{\ref@jnl{Fund.~Cosmic~Phys.}}  
\def\gca{\ref@jnl{Geochim.~Cosmochim.~Acta}}   
\def\grl{\ref@jnl{Geophys.~Res.~Lett.}} 
\def\jcp{\ref@jnl{J.~Chem.~Phys.}}      
\def\jgr{\ref@jnl{J.~Geophys.~Res.}}    
\def\jqsrt{\ref@jnl{J.~Quant.~Spec.~Radiat.~Transf.}}
\def\memsai{\ref@jnl{Mem.~Soc.~Astron.~Italiana}}
\def\nphysa{\ref@jnl{Nucl.~Phys.~A}}   
\def\physrep{\ref@jnl{Phys.~Rep.}}   
\def\physscr{\ref@jnl{Phys.~Scr}}   
\def\planss{\ref@jnl{Planet.~Space~Sci.}}   
\def\procspie{\ref@jnl{Proc.~SPIE}}   

\makeatother

\newcommand{\nus}{\emph{NuSTAR}}

\newcommand{\rxte}{\,\emph{RXTE}}

\newcommand{\swift}{\emph{Swift}}

\newcommand{\suzaku}{\,\emph{Suzaku}}

\newcommand{\cgro}{\,\emph{CGRO}}

\newcommand{\xrt}{\,\emph{XRT}}
\newcommand{\fpm}{\,\emph{FPM}}
\newcommand{\arielfive}{\,\emph{Ariel$-$5}}

\newcommand{\xray}{X-ray}

\newcommand{\src}{1A\,1118}
\newcommand{\source}{1A\,1118-61}

\newcommand{\psec}{s$^{-1}$}  
 
\begin{document}

\title{Accretion geometry and spectral evolution in \source: a comparison of the 2009 and 2026 outbursts}

\author[0000-0002-7391-5776]{Kinjal Roy}
\affiliation{Raman Research Institute, C. V. Raman Avenue, Sadashivanagar, Bengaluru - 560 080, India.}
\email[show]{kinjal@rrimail.rri.res.in} 

\author[0000-0003-3753-3102]{Aru Beri}
\affiliation{Indian Insitiute of Astrophysics, 2nd Block, Koramangala, Bangalore-560 034, India.}
\affiliation{School of Physics and Astronomy, University of Southampton, Southampton, SO17 1BJ, UK.}
\email[show]{aru.beri@iiap.res.in}

\author[0000-0003-0366-047X]{Rahul Sharma}
\affiliation{Inter-University Centre for Astronomy and Astrophysics, Ganeshkhind, Pune, 411007, India.}
\email[]{}

\author[]{Phil Charles}
\affiliation{School of Physics and Astronomy, University of Southampton, Southampton, SO17 1BJ, UK.}
\email[]{}


\begin{abstract}

We present a detailed spectro-temporal study of the Be X-ray binary pulsar \source\ during its brightest recorded outburst in 2026, using \swift\ and \nus\ observations, and compare its properties with the 2009 outburst. Coherent pulsations at $\sim400$ s are detected throughout the outburst, with pulse profiles evolving strongly with energy and luminosity, indicating changes in emission geometry. A transient quasi-periodic oscillation (QPO) at $\sim$0.11 Hz is observed during the rising phase. The luminosity dependence of the QPO frequency during the current and previous outbursts suggests an origin associated with instabilities near the magnetospheric radius. 
The broadband spectra are well described by thermal Comptonization and show clear spectral hardening at higher luminosities. A cyclotron line is detected in the two \nus\ observations, with its energy remaining nearly constant despite a factor of $\sim25$ change in luminosity. Long-term monitoring reveals that the 2026 outburst is systematically harder and brighter, suggesting significant difference in the accretion structure and emission regions between the two outbursts.

\end{abstract}

\keywords{\uat{High Energy astrophysics}{739} --- \uat{X-ray binary stars}{1811} --- \uat{High mass x-ray binary stars}{733} --- \uat{X-ray astronomy}{1810} --- \uat{Neutron stars}{1108}}


\section{Introduction}

Transient X-ray binary pulsars with Be companions (BeXRP) are uniquely suited to test accretion onto highly magnetized neutron stars (NSs). The BeXRPs are known to exhibit a wide range of luminosity, ranging from $10^{37-38}$ erg s$^{-1}$ or even up to  $10^{39}$ erg s$^{-1}$ during the peak of outbursts to $10^{32-34}$ erg s$^{-1}$ during the quiescence phase~\citep{Be_transient_Tsygankova_2017}. Pioneered by the highly successful \textit{RXTE}-ASM~\citep{RXTE_ASM}, the practice of initiating observations soon after the onset of an outburst has become standard. Recent advances in sensitive, wide-viewing-angle X-ray instruments such as \textit{MAXI}/GSC~\citep{maxi_main_paper}, \textit{Swift}/BAT~\citep{SWIFT_BAT}, and \textit{Fermi}/GBM~\citep{Fermi_GBM} enable the prompt identification of transients entering an outburst. Subsequent follow-up observations with sensitive broadband instruments such as \textit{Nuclear Spectroscopic Telescope Array}~\citep[\nus;][]{NuSTAR_Harrison_2013} and \textit{Neil Gehrel’s Swift Observatory}~\citep[\swift;][]{Swift_XRT_Burrows_2005} enable comprehensive studies of the source's characteristics across a wide range of energies.

The transient X-ray source \source\ (henceforth \src) was serendipitously discovered during an observation of the nearby X-ray pulsar Cen X-3, when an outburst was detected with the \arielfive\ satellite in December 1974~\citep{1A118_discovery_Eyles_1975}. Since then, it has exhibited only a handful of major outbursts. A second outburst occurred in January 1992 when \cgro/BATSE (20-100 keV energy range) recorded a peak flux of ~150 mCrab \citep{1A1118_1992_outbust_Coe_1994}, similar to that in the 1974 outburst. A third outburst occurred in 2009, when the source was detected by \swift/BAT, and followed up with \suzaku\ and \rxte\ \citep{1A1118_Doroshenko_RXTE_2010, 1A1118_RXTE_PCA_Jincy_2011, 1A1118_Suzaku_Chandreyee_2012}.  And the most recent outburst was detected in 2026, when it reached its brightest recorded state ~\citep{Swift_2026_outburst_ATel}. Coherent X-ray pulsations with a period of $\sim405.6$ s were identified early on~\citep{1A1118_Ives_1975, 1A1118_Fabian_1975}. The optical counterpart ‘WRA 793’ is a highly reddened Be star classified as an O9.5IV–Ve star~\citep{1A1118_companion_Chevalier_1975}. The source distance was estimated to be $2.9\pm0.1$ kpc from \textit{GAIA} DR3 data ~\citep{GAIA_DR3_2021, GAIA_DR3_2023}.

In these systems, matter from the companion star is funneled by the strong field lines   ($B \gtrsim 10^{12}$ G) onto the NS's magnetic poles. The resulting quantization of electron energy levels gives rise to cyclotron resonant scattering features (CRSFs) in the X-ray spectrum, enabling direct measurement of the magnetic field strength~\citep{Meszaros_1992_book}. Despite their diagnostic importance, CRSFs have been detected in only a fraction of known accreting pulsars~\footnote{\url{http://orma.iasfbo.inaf.it:7007/~mauro/pulsar_list.html}}, and their presence remains not fully understood.

The properties of CRSFs are known to vary with source luminosity, providing insight into the structure of the accretion column~\citep{CRSF_review_Staubert_2019}. In particular, the cyclotron line energy can exhibit either a positive or negative correlation with luminosity, depending on whether the source is in the sub-critical (Coulombic shock regime) or super-critical (radiation-dominated) accretion regime. Transitions between these regimes have been observed in sources such as A 0535$+$26~\citep{A0535p26_Kong_2021}, V 0332$+$53~\citep{V0332p53_CRSFvslumin_Lutovinov_2015, V0332p53_Doroshenko_2017} and 4U 0115+63~\citep{4U0115p63_Roy_2024}, and are sometimes accompanied by changes in pulse profile morphology~\citep{V0332p53_CRSFvslumin_Lutovinov_2015, Critical_lumin_WilsonHodge_2018, Beri2020, RX_J0440p9p4431_QPO_Rahul_2024}. In contrast, some systems, such as GX 304$-$1~\citep{GX304m1_CRSF_Rothschild_2017} and Cep X$-$4~\citep{CepX4_CRSF_Roy_2025}, show a saturation of cyclotron line energy at high luminosities, suggesting more complex accretion geometries.

Quasi-periodic oscillations (QPOs) provide an additional probe of the inner accretion flow. They are generally attributed to inhomogeneities or instabilities in the accretion disc, particularly near the magnetospheric boundary~\citep{QPO_review_Finger_1998, KS_1947p300_QPO_Marykutty_2010, QPO_pulsars_Hemanth_2024}. QPOs are typically exhibited by accreting high magnetic field X-ray pulsars at frequencies of a few tens of mHz. QPOs in \src\ have been detected with \rxte\ during the 2009 outburst, ranging from  $70-90$ mHz~\citep{1A1118_RXTE_PCA_Jincy_2011}. Observations with \suzaku\ also showed that the QPO frequency varied with Luminosity according to $\nu_{QPO} \propto L_X^{3/7}$~\citep{1A1118_Suzaku_Chandreyee_2012}.

The 2026 outburst of \src, the brightest observed to date, offers a unique opportunity to investigate accretion dynamics and magnetic field properties of the NS over an extended luminosity range. In this work, we present a comprehensive timing and spectral study of \src\ using coordinated observations from \swift\ (XRT) and \nus. The \swift\ data enable us to track the long-term evolution of the outburst and probe the soft X-ray regime, while \nus\ provides sensitive broadband ($3–79$ keV) coverage crucial for detailed spectral modeling, including cyclotron line. We further perform a comparative analysis of the 2009 and 2026 outbursts to examine the evolution of spectral and timing properties across different luminosity states. We investigate the dependence of the cyclotron line energy on luminosity and explore its implications for accretion geometry and shock structure in highly magnetized NSs.

The paper is structured as follows. In Section~\ref{sec: inst}, we describe the data reduction and analysis procedures. In Section~\ref{sec:analysis}, we present the timing and spectral results, including phase-resolved and time-resolved studies. In Section~\ref{sec: disc}, we discuss the implications of our findings in the context of accretion physics and compare the characteristics of the two outbursts.

\section{Observations and Data Reduction} \label{sec: inst}

\begin{figure}
    \centering
    \includegraphics[height=0.75\linewidth,width=\linewidth]{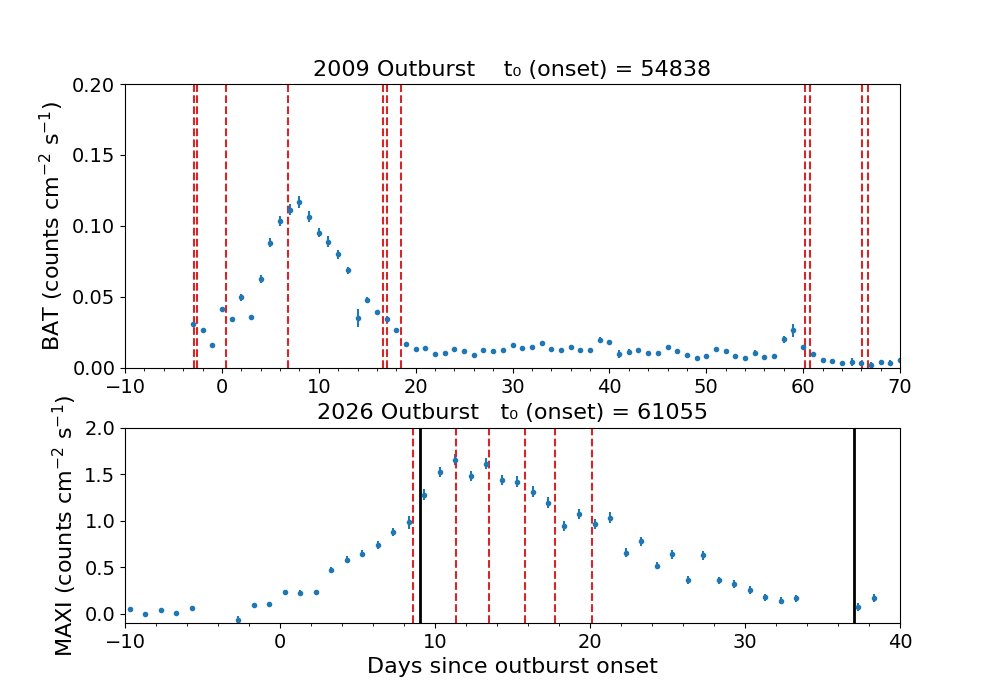}
    \caption{2009 and 2026 outburst of \src\ as seen by \swift/BAT and \textrm{MAXI}/GSC, respectively. Top panel shows the light curve during the 2009 outburst from \swift/BAT in the 15-50 keV range. Bottom panel shows the 2-20 keV \textrm{MAXI}/GSC count rate of the source during the 2026 outburst. Red dotted line marks the \swift/XRT observations. The two vertical lines in black are the two \textit{NuSTAR} observations of the source carried out during the 2026 outburst. }
    \label{fig:BAT-MAXI-outburst}
\end{figure}

The outburst light curve of \src\ during the 2009 (top panel) and 2026 (bottom panel) outbursts is shown in Fig.~\ref{fig:BAT-MAXI-outburst}. The light curve for the 2026 outburst was obtained from \texttt{MAXI}~\footnote{\url{http://maxi.riken.jp/mxondem/}}~\citep{maxi_main_paper} in the 2$-$20 keV energy range, binned to intervals of one day. For the 2009 outburst, we used the long-term \textit{Swift}/BAT light curve~\citep{SWIFT_BAT} in the 15$-$50 keV energy range, also binned to one-day intervals. The epoch of \nus\ observations during the 2026 outburst are marked with black vertical lines, while the \swift/\xrt\ observations are marked with red dashed lines in Fig.~\ref{fig:BAT-MAXI-outburst}. Observation details are provided in Table \ref{tab:obs_detail}.

\begin{table*}
    \centering
    \caption{Observational details of \src's 2026 and 2009 outbursts.} 
    \begin{tabular}{|c|c|c|c|c|}
        \hline
        \multicolumn{5}{|c|}{\textbf{2026 Outburst}} \\ \hline
        Observation ID & Instrument & MJD & Exposure~(ks) & $P_{spin}$ (sec) \\ \hline
        
        00031370006 (Obs 1) & \swift-\xrt\ & 61063 & 1 & $409.5\pm0.5$  \\
        00031370007 (Obs 2) & \swift-\xrt\ & 61066 & 1   & $408.7\pm0.5$ \\ 
        00031370008 (Obs 3) & \swift-\xrt\ & 61068 & 0.5 & $408^{a}$  \\   
        00031370009 (Obs 4) & \swift-\xrt\ & 61070 & 1   & $407.5\pm0.5$\\
        01442056001 (Obs 5) & \swift-\xrt\ & 61072 & 1   & $409^{a}$ \\ 
        01442056002 (Obs 6) & \swift-\xrt\ & 61075 & 1   & $408^{a}$ \\ 
        91201305002 (NuS-1) & \nus-\fpm\ & 61064 & 14    & $408.77\pm0.03$  \\ 
        91201310002 (NuS-2) & \nus-\fpm\ & 61092 & 28    & $408.16\pm0.01$ \\ \hline

        \multicolumn{5}{|c|}{\textbf{2009 Outburst }} \\ \hline
        
        00339131000~(XRT 1) & \swift-\xrt\ & 54835 & 5.0     & $406^{a}$ \\ 
        00339131001~(XRT 2) & \swift-\xrt\ & 54835 & 0.5     & $406^{a}$ \\   
        00339131002~(XRT 3) & \swift-\xrt\ & 54838 & 2.7     & ${407.2\pm0.1}^b$ \\ 
        00339131003~(XRT 4) & \swift-\xrt\ & 54838 & 2.7    & ${407.2\pm0.1}^b$ \\
        00339853000~(XRT 5) & \swift-\xrt\ & 54844 & 1.6     & $403^{a}$ \\ 
        00340919000~(XRT 6) & \swift-\xrt\ & 54854 & 1.4     & ${407.5\pm0.5}^c$ \\ 
        00340973000~(XRT 7) & \swift-\xrt\ & 54855 & 3.0     & ${407.5\pm0.5}^c$ \\ 
        00341076000~(XRT 8) & \swift-\xrt\ & 54856 & 1.0     & ${407.5\pm0.5}^c$ \\
        00031370001~(XRT 9) & \swift-\xrt\ & 54894 & 4.2     & ${407.3\pm0.1}^d$  \\ 
        00031370002~(XRT 10) & \swift-\xrt\ & 54898 & 5.1     & ${407.3\pm0.1}^d$ \\ 
        00031370003~(XRT 11) & \swift-\xrt\ & 54898 & 3.2    & ${407.3\pm0.1}^d$ \\ 
        00031370004~(XRT 12) & \swift-\xrt\ & 54904 & 5.6      & ${407.1\pm0.5}^e$ \\
        00031370005~(XRT 13) & \swift-\xrt\ & 54904 & 4.1    & ${407.1\pm0.5}^e$  \\ \hline
    \end{tabular}
    \footnote{: Period estimates from \texttt{efsearch} with low S/N; not statistically significant. $^{b, c, d, e}$: Combined observations used to improve S/N for period detection.}
    \label{tab:obs_detail}
\end{table*}

\subsection{\nus}
\nus\ is a broadband X-ray imaging and spectroscopic mission. \nus\ has hard \xray\ focusing optics with two identical co-aligned focal plane modules (\fpm), namely FPMA and FPMB~\citep{NuSTAR_Harrison_2013}. \nus\ has a broadband energy coverage of 3$-$79 keV, with a spectral resolution of 400 eV at 10 keV. \nus\ made two observations of \src, four weeks apart on 24th Jan and 21st Feb 2026, see Table~\ref{tab:obs_detail} for details.

The \nus\ data was reduced using \texttt{HEASOFT} v6.36 along with the latest calibration files available via \texttt{CALDB} version 20251215 for \nus/\fpm. The \texttt{nupipeline} version 0.4.9 was used to produce the clean event files. 
For the first \nus\ observation, we processed the data with additional keyword \texttt{statusexpr="(STATUS$==$b0000xxx00xxxx000) \&\& (SHIELD$==$0)"} since the source count rate exceeds 100 counts \psec.
The \texttt{nuproduct} command was used to extract the light curves and spectrum corresponding to the source and background regions. A 125\arcsec\ diameter circle around the source position was used for extracting the source level 2 products. Similarly, a circle of the same size, located away from the source region, was used to extract the background products. Barycenter correction was performed using \texttt{barycorr} v2.16. 

\subsection{\swift}
\swift, one of NASA's Medium Explorer (MIDEX) missions for monitoring of GRBs, was launched in 2004. There are three scientific instruments onboard \swift\ observatory operating over a wide energy range. The three payloads are the X-Ray Telescope~\citep[XRT -- 0.2-10 keV;][]{Swift_XRT_Burrows_2005}, Burst Alert Telescope~\citep[BAT -- 15-150 keV;][]{Swift_BAT_Barthelmy_2005}, and the Ultraviolet-Optical Telescope~\citep[UVOT -- 70-600 nm;][]{Swift_UVOT_Roming_2004}. The \swift\ observatory was designed for prompt follow-up observations of highly energetic GRBs and is capable of rapidly changing its pointing direction to observe new transient sources. In this work, we use the multiple observations of \swift/XRT during the 2009 and 2026 outbursts, the details of which are given in Table~\ref{tab:obs_detail}. One of these observations (Obs 1) was made simultaneously with the first \nus\ observation of the 2026 outburst. We have used data collected in the \textit{Windowed Timing} (WT) mode for all observations.

The data reduction was performed using the standard data pipeline package \texttt{XRTPIPELINE} version \texttt{0.13.7} in order to produce cleaned event files using the latest calibration files (version 20250609). The \swift/\xrt\ observations were taken in the WT mode. The source events were extracted using a circular aperture of radius 60\arcsec\, and the background events were extracted from the same size aperture away from the source, and free of any other sources. The source count rate was below the pile-up limit of $\sim$100 counts/s, so no pile-up corrections were performed for the data. The spectra were obtained from the corresponding event files using the \texttt{XSELECT} tool, and the ancillary file was created using the tool \texttt{xrtmkarf}. The RMF file is sourced from the calibration database (version 20250609).

\section{Analysis} \label{sec:analysis}

\subsection{Timing Analysis} \label{sec: timing analysis}

\begin{figure}
    \centering
    \includegraphics[width=0.95\linewidth]{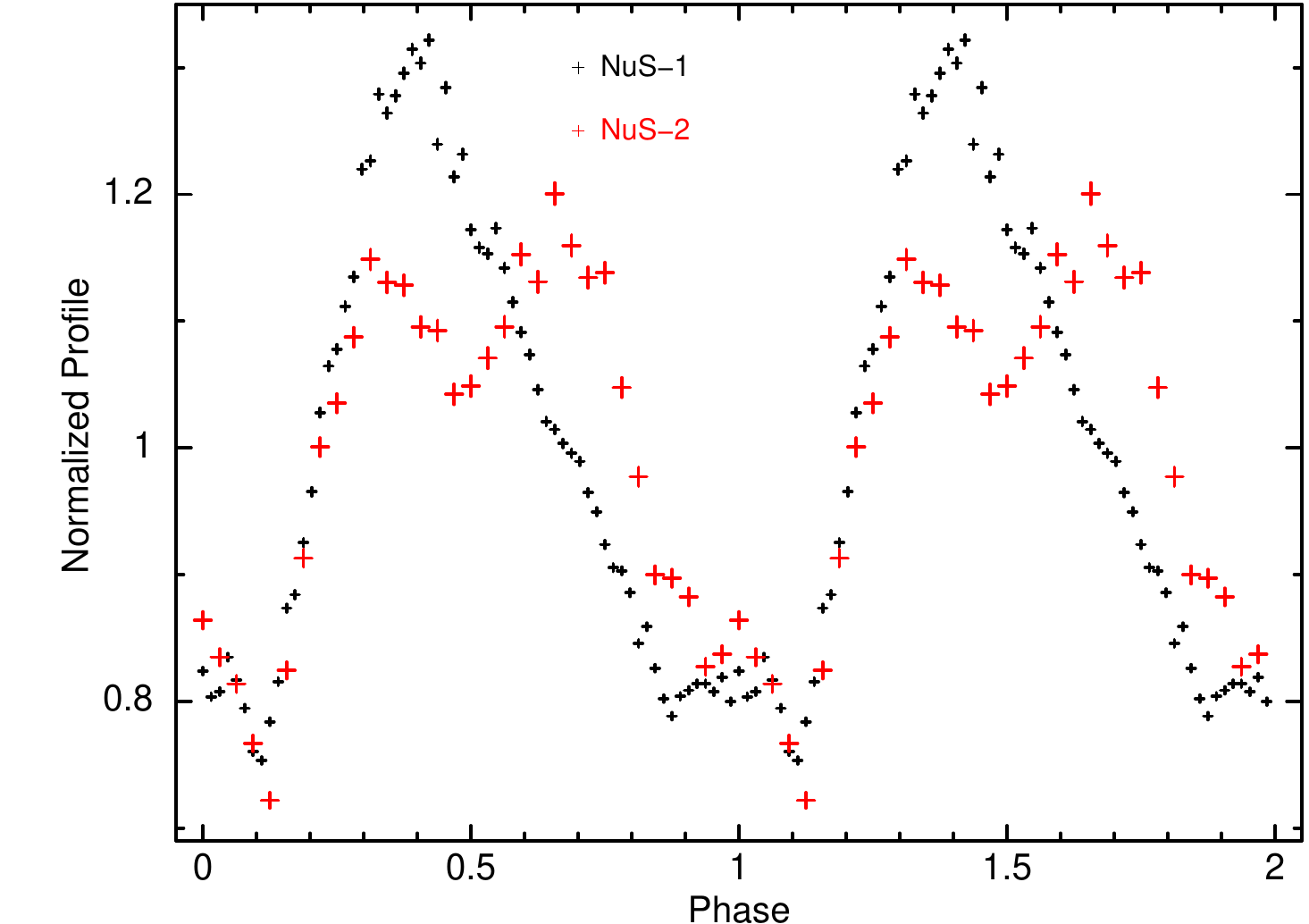}
    \caption{Pulse profiles of \src\ from the two \nus\ observations in the $3$–$79$ keV energy range. The first observation is shown in black, while the second observation is shown in red.}
    \label{fig:folded profile NuStar}
\end{figure}

Light curves from the \nus\ observations were extracted with a time resolution of 1 s. The first \nus\ observation was taken during the rising phase of the 2026 outburst (Fig.~\ref{fig:BAT-MAXI-outburst}) and shows a combined count rate of $\sim$570 counts s$^{-1}$ from the two \nus\ FPM modules. \src\ is known to exhibit pulsations with a period of $\sim 406$ s. Using the epoch folding~\citep{Leahy_efsearch_1987} tool \texttt{efsearch}, we determined the spin period to be 408.77 (3) s. The corresponding folded profile is shown in the top panel of Fig.~\ref{fig:folded profile NuStar}, primarily composed of a strong single peak. 

The second \nus\ observation (NuS-2), obtained toward the end of the outburst, shows a significantly lower count rate of $\sim$27 counts s$^{-1}$ (Fig.~\ref{fig:BAT-MAXI-outburst}). Pulsations remain detectable, with a measured period of 408.16(1) sec. The pulse profile evolved markedly, changing from a single-peaked structure in the first observation to a double-humped shape (shown in red in Fig.~\ref{fig:folded profile NuStar}. For consistency, the profiles from both epochs are phase-aligned such that the minimum occurs near pulse phase $\sim$0.1. 

We also performed timing analysis using the WT mode data of \swift/XRT during the 2009 and 2026 outbursts. The \xrt\ light curves were extracted with a time resolution of 1~\rm{s}. Significant pulsations were detected in a subset of the observations, and the corresponding spin periods are reported in Table~\ref{tab:obs_detail}. For some observations (marked as `a' in Table~\ref{tab:obs_detail}), the signal-to-noise ratio was insufficient for statistically significant detections; however, approximate period values inferred from \texttt{efsearch} are listed for completeness. These observations were therefore excluded from performing pulse profile analysis. During the 2009 outburst, the observations (marked as b,c,d,e) were combined to increase the signal-to-noise ratio and period detection significance. The corresponding combined observations were used to construct pulse profiles. \\

\begin{figure}
    \centering
    \includegraphics[width=0.48\linewidth]{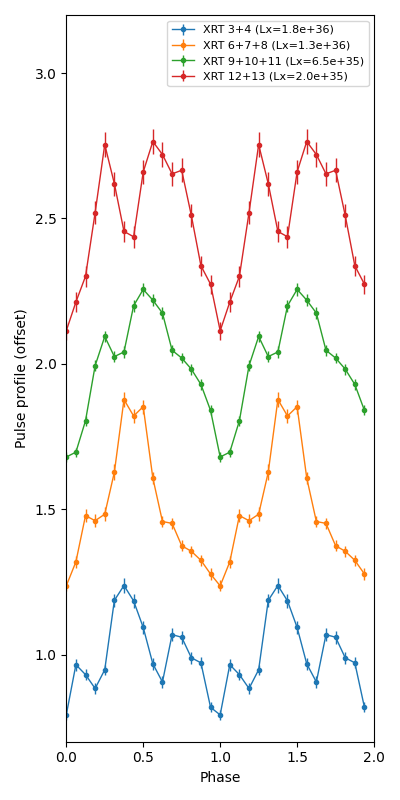}
    \includegraphics[width=0.48\linewidth]{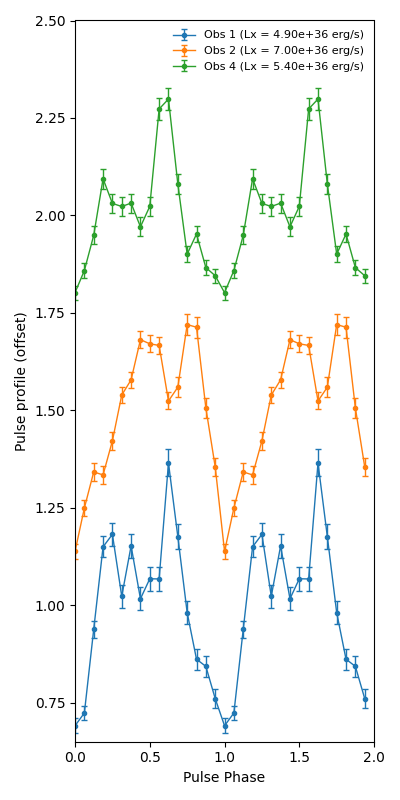}
    \caption{Evolution of \xrt\ pulse profiles of \source\ during 2009 outburst ({\it Left}) and 2026 outburst ({\it Right}). The profiles were phase-shifted so as to align the dip features.}
    \label{fig:efold-XRT}
\end{figure}

Fig.~\ref{fig:efold-XRT} shows the evolution of the pulse profiles of \src\ obtained from \swift/XRT observations in the $0.5–10$ keV energy band during the 2009 and 2026 outbursts. The evolution of the \swift/XRT pulse profiles shows distinct behavior during the two outbursts. During the 2026 outburst, the pulse profiles exhibit a complex, double-peaked structure. In contrast, during the 2009 outburst, the brightest observation ($\sim 1.8 \times 10^{36}$ erg s$^{-1}$) exhibits a multi-peaked structure that evolves into simpler, single- and double-peaked profiles at lower luminosities. 
Due to the limited exposure and relatively limited statistics of individual \xrt\ observations, we do not compute pulse fractions. Instead, we focus on the qualitative evolution of the pulse profiles.

The pulse fraction (PF) is used to measure the variability of a pulse profile, and is defined as:
\begin{equation}
    PF_{RMS} = \frac{1}{\overline{p} \sqrt{N}} \sqrt{\sum_{i=1} ^{N} ( (p_i - \overline{p})^2 - \sigma_i^2 ) } 
\end{equation}
where $\overline{p}$ is the average pulse profile, $p_i$ and $\sigma_i$ are the pulse profile rate and error in $i^{th}$ phase bin, and N is the total number of phase bins. The PF in the $3-79$ keV range was calculated to be $ 17.86 \pm 0.01 $ \% during the first \nus\ observation (NuS-1), decreasing to $13.48 \pm 0.03$ \% in the second observation (NuS-2). 

\subsubsection{Pulse Energy Dependence}

\begin{figure}
    \centering
    \includegraphics[width=0.45\linewidth]{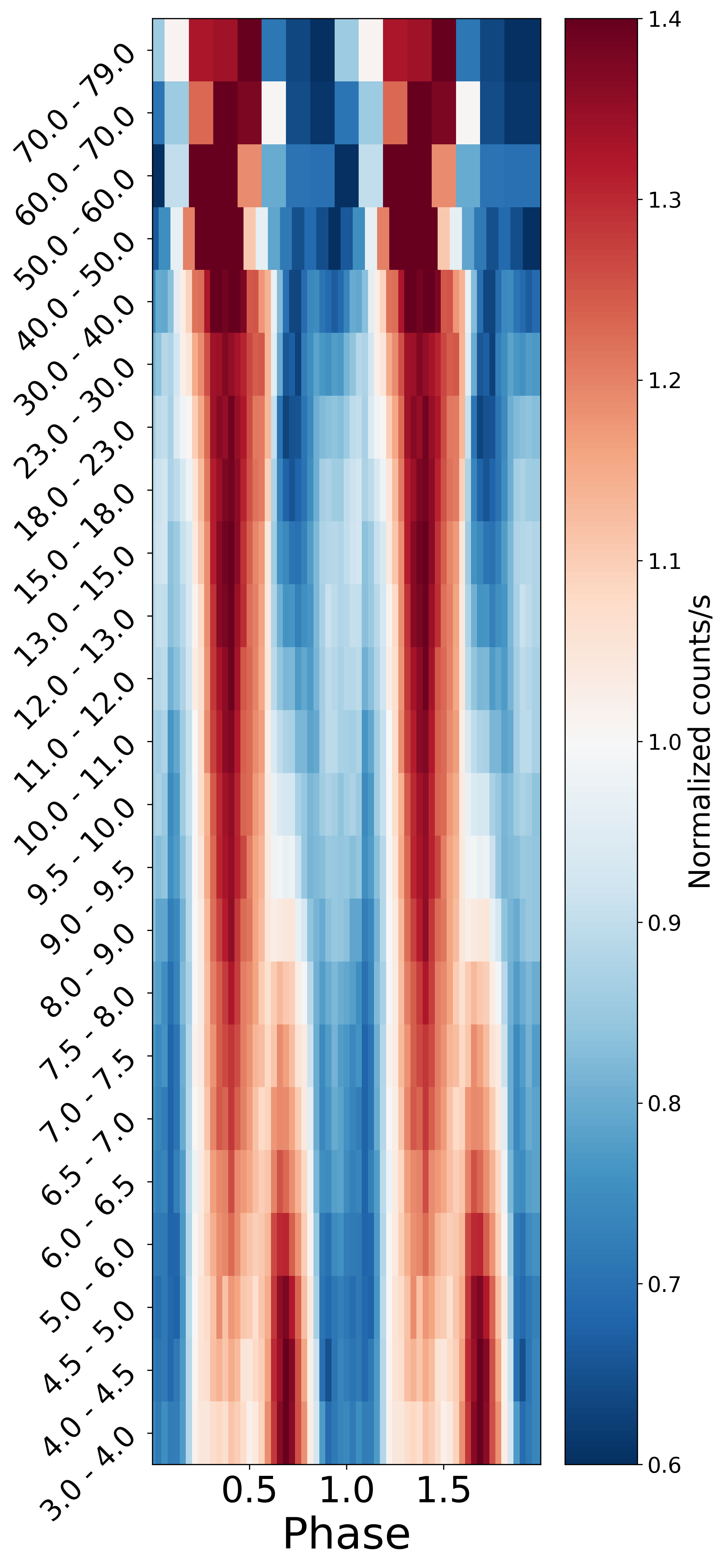}
    \includegraphics[width=0.45\linewidth]{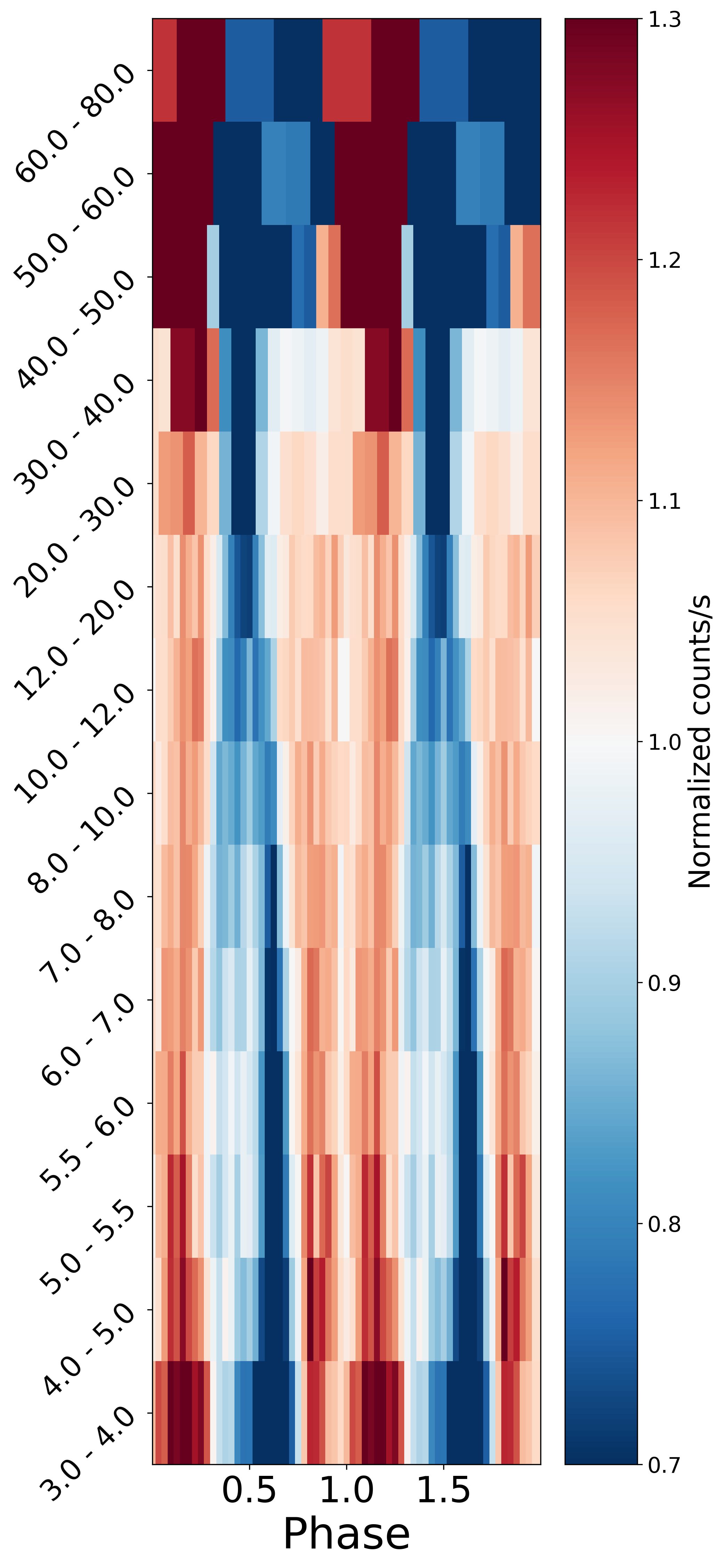}
    \caption{Energy-resolved (in 23 bands, given in keV) pulse profiles of \source\ from the two NuSTAR observations NuS-1 ({\it Left}) and NuS-2 ({\it Right}). The colour bar shows the normalised count rate from high (red) to low (blue)}
    \label{fig:energy_dependent_PP}
\end{figure}

\begin{figure}
    \centering
    \includegraphics[width=0.95\linewidth]{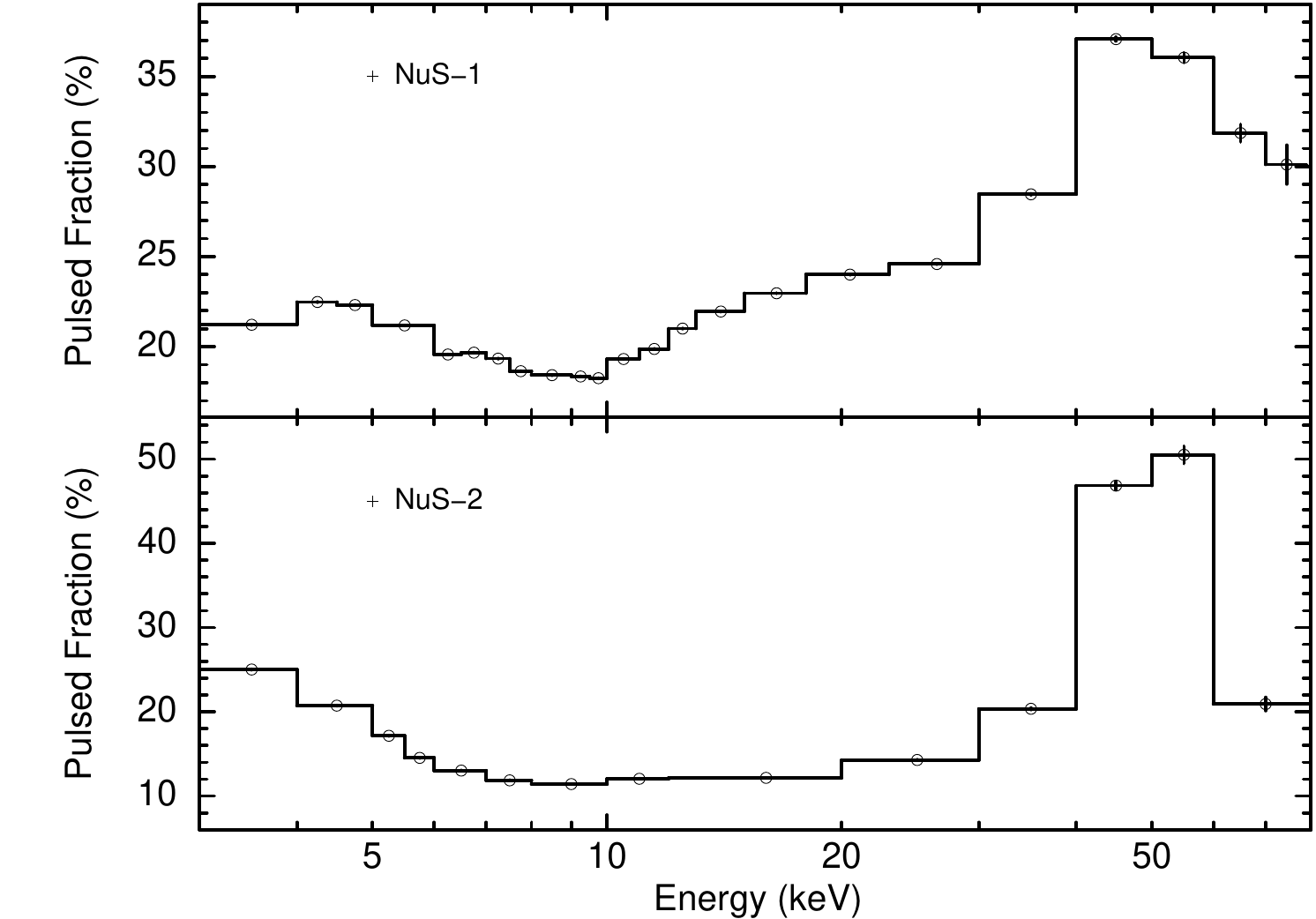}
    \caption{Evolution of the pulsed fraction with energy for the two \nus\ observations of \src.}
    \label{fig:energy dependent pf}
\end{figure}

The full \nus\ energy range was divided into multiple sub-bands to investigate the energy dependence of the timing properties. Energy-resolved pulse profiles for NuS-1 are shown in the left panel of Fig.~\ref{fig:energy_dependent_PP}. At lower energies, the profile is dual-peaked with the prominent peak near phase $\sim0.7$. With increasing energy, a secondary peak emerges around phase $\sim0.4$, and the profile gradually evolves toward a more single-peaked morphology, accompanied by an increase in modulation amplitude. The corresponding pulsed fraction as a function of energy is shown in the top panel of Fig.~\ref{fig:energy dependent pf}, revealing a clear increase with energy.

A similar analysis was carried out for NuS-2. The energy-resolved profiles (right panel of Fig.~\ref{fig:energy_dependent_PP}) show complex, multi-peaked structures with a pronounced trough at lower energies. As energy increases, the profiles evolve into a single-peaked shape, with the phase of the trough drifting with energy. The pulsed fraction (bottom panel of Fig.~\ref{fig:energy dependent pf}) also increases with energy, reaching a maximum around $\sim50$ keV before declining at higher energies.~This trend was observed in both Nus-1 and Nus-2.

\subsubsection{Power Density Spectrum}
\begin{figure}
    \centering
    \includegraphics[width=0.95\linewidth]{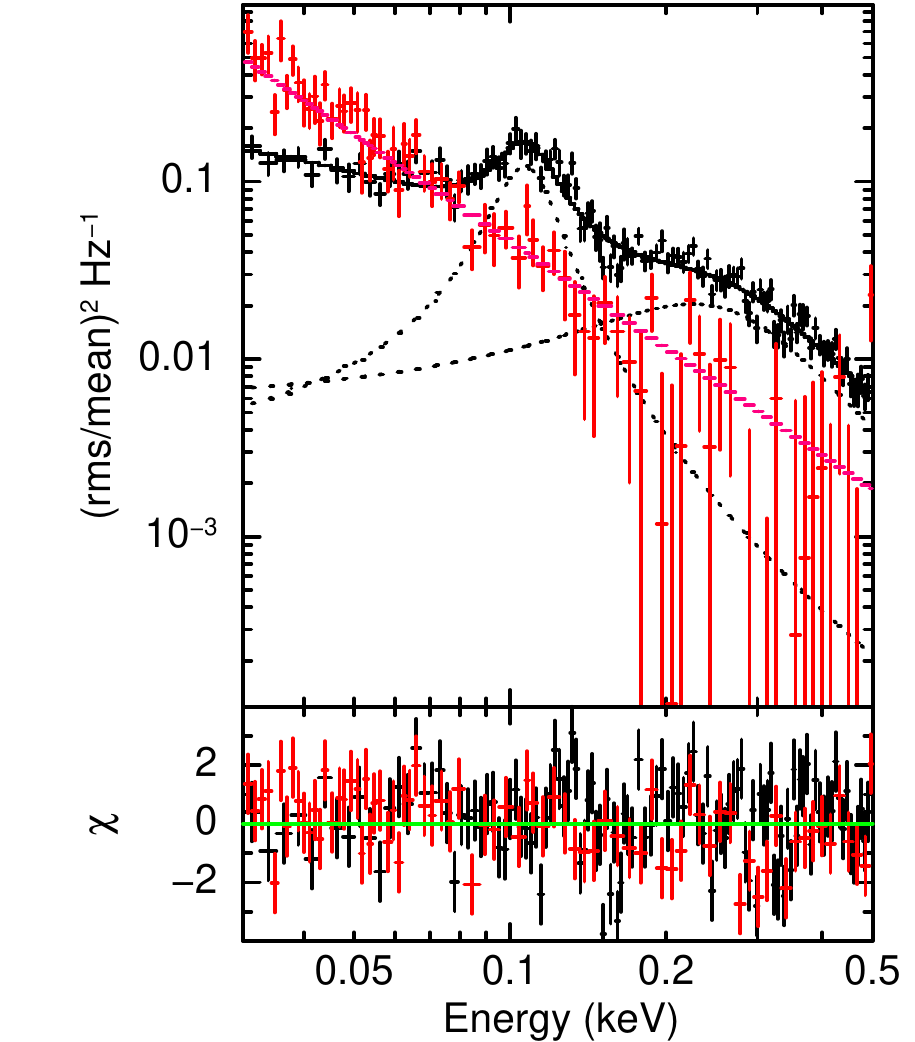}
    \caption{({\it Top}) Power density spectrum (3-79keV) of \src\ during NuS-1 (black) and NuS-2 (red), together with the model components. {\it Bottom}) Residuals between the data and best-fit model.}
    \label{fig:power spectrum}
\end{figure}

We extracted the power density spectrum (PDS) from the two \nus\ observations. The power spectra were normalized such that their integral gives the squared rms fractional variability, and the expected white-noise level was subtracted. The resultant PDS is shown in Fig.~\ref{fig:power spectrum} in black for NuS-1 and in red for NuS-2. The PDS is shown over the frequency range of $0.03-0.5$ Hz, to avoid contamination from the spin frequency and its harmonics. 

A prominent feature is detected at $\sim$0.1 Hz in the first \nus\ observation. The broadband PDS is well modeled with multiple \texttt{Lorentzian} profiles. The \texttt{Lorentzian} corresponding to the QPOs is at $110_{-10}^{+10}$ mHz, with a quality factor of $3.3_{-0.9}^{+0.9}$ and rms of $8.0^{+0.7}_{-0.8}$\%. A secondary feature is also present along with the main QPO in the \nus\ PDS at about 0.22 Hz. The best-fit parameters of the QPO components are listed in Table~\ref{tab:qpo_details}. In contrast, the PDS from the NuS-2, plotted in red in Fig.~\ref{fig:power spectrum}, does not exhibit any significant peak near the QPO frequency. The PDS of the NuS-2 could be well modeled with a zero-center \texttt{Lorentzian}, the residuals of which are plotted in red in the bottom panel of Fig.~\ref{fig:power spectrum}. Assuming that the QPO is present at the frequency observed during the first \nus\ spectrum we get an upper limit on its r.m.s. value of 2.4\%.

\begin{figure}
    \centering
    \includegraphics[width=0.95 \linewidth]{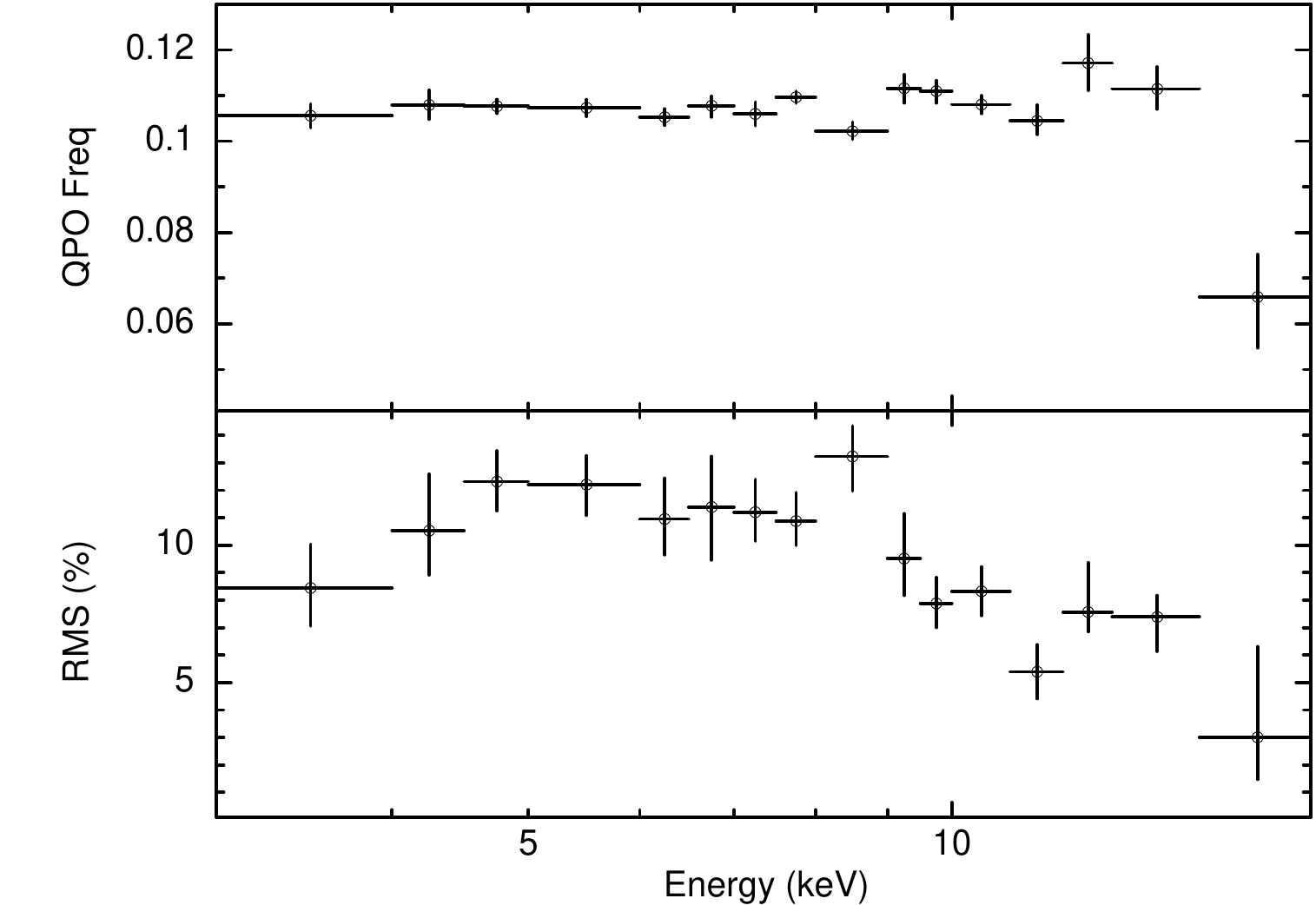}
    \caption{Energy dependence of the QPOs parameters as observed in NuS-1.}
    \label{fig:QPO energy evolution}
\end{figure}

The QPO is detected across most energy bands below $\sim$18 keV. To investigate its energy dependence, we fitted the same model to energy-resolved PDS. The energy dependence of the primary QPO parameters is shown in Fig.~\ref{fig:QPO energy evolution}. The Lorentzian width corresponding to the QPO could not be constrained in the individual energy-resolved PDS and was fixed to the broadband fit value.

The QPO frequency and fractional rms both exhibit noticeable energy dependence. The rms increases with energy up to 5 keV, remains constant around $\sim5-9$ keV, and then declines, with no significant detection above $\sim$18 keV. The QPO frequency remains approximately constant at higher energies (up to 15 keV) and decreases to $\sim$0.06 Hz in the $15$–$18$ keV range (top panel of Fig.~\ref{fig:QPO energy evolution}). This behavior is consistent with that reported during the 2009 outburst using \rxte\ \citep{1A1118_RXTE_PCA_Jincy_2011}.

\begin{table}
    \centering
    \caption{Details of \src\ quasi-periodic oscillations.}
    \begin{tabular}{|c|c|}
        \hline
        Parameter             & $\nu_{QPO}$ \\ \hline
        Frequency (Hz)       & $0.11_{-0.01}^{+0.01}$  \\ \hline
        Width (Hz)           & $0.033_{-0.006}^{+0.006}$  \\ \hline
        Quality Factor        & $3.3_{-0.9}^{+0.9}$  \\ \hline
        Fractional r.m.s (\%) & $8.0_{-0.8}^{+0.7}$  \\ \hline
    \end{tabular}
    \label{tab:qpo_details}
\end{table}

\subsection{Spectral Analysis} \label{sec: sp analysis}

During NuS-1, simultaneous observations of \src\ with \swift\ and \nus\ were used for broadband spectral analysis. To jointly fit the spectra from \swift/\xrt\ and \nus/\fpm, a cross-normalization factor was included, which was fixed to unity for \nus\ FPMA and allowed to vary for the \xrt\ and \nus\ FPMB. All the spectral analysis was performed using \texttt{XSPEC} version \texttt{12.12.1}~\citep{XSPEC} with the interstellar absorption modelled using the Tuebingen-Boulder absorption model \texttt{TBabs}, with abundances from~\citet{Wilm_abund} and photoelectric cross-sections from~\citet{Vern_cs}. The spectra were optimally binned based on~\citet{Optimal_binning_Kaastra_Bleeker_2016}.

\begin{figure*}
    \centering
    \includegraphics[width=0.49\linewidth]{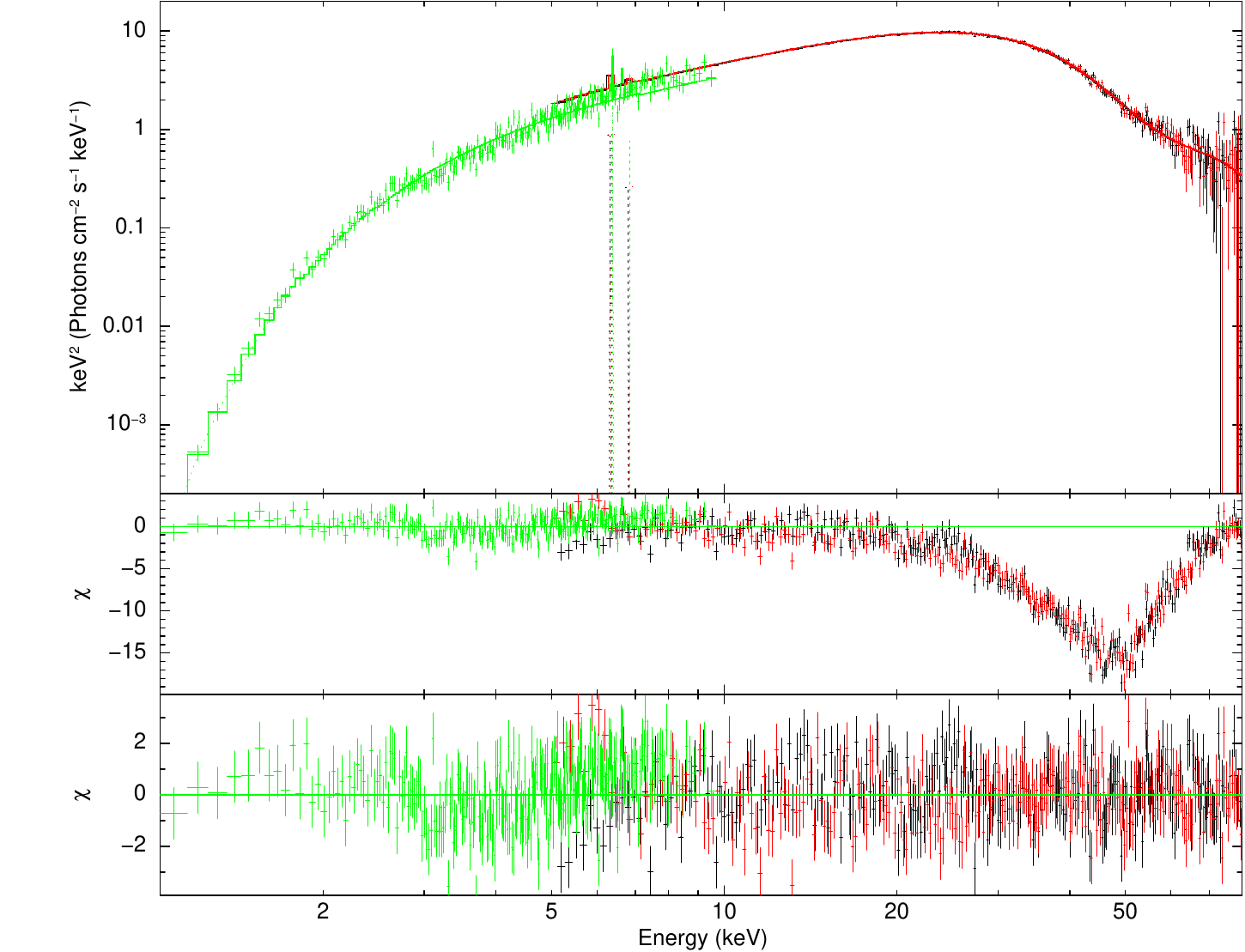}
    \includegraphics[width=0.49\linewidth]{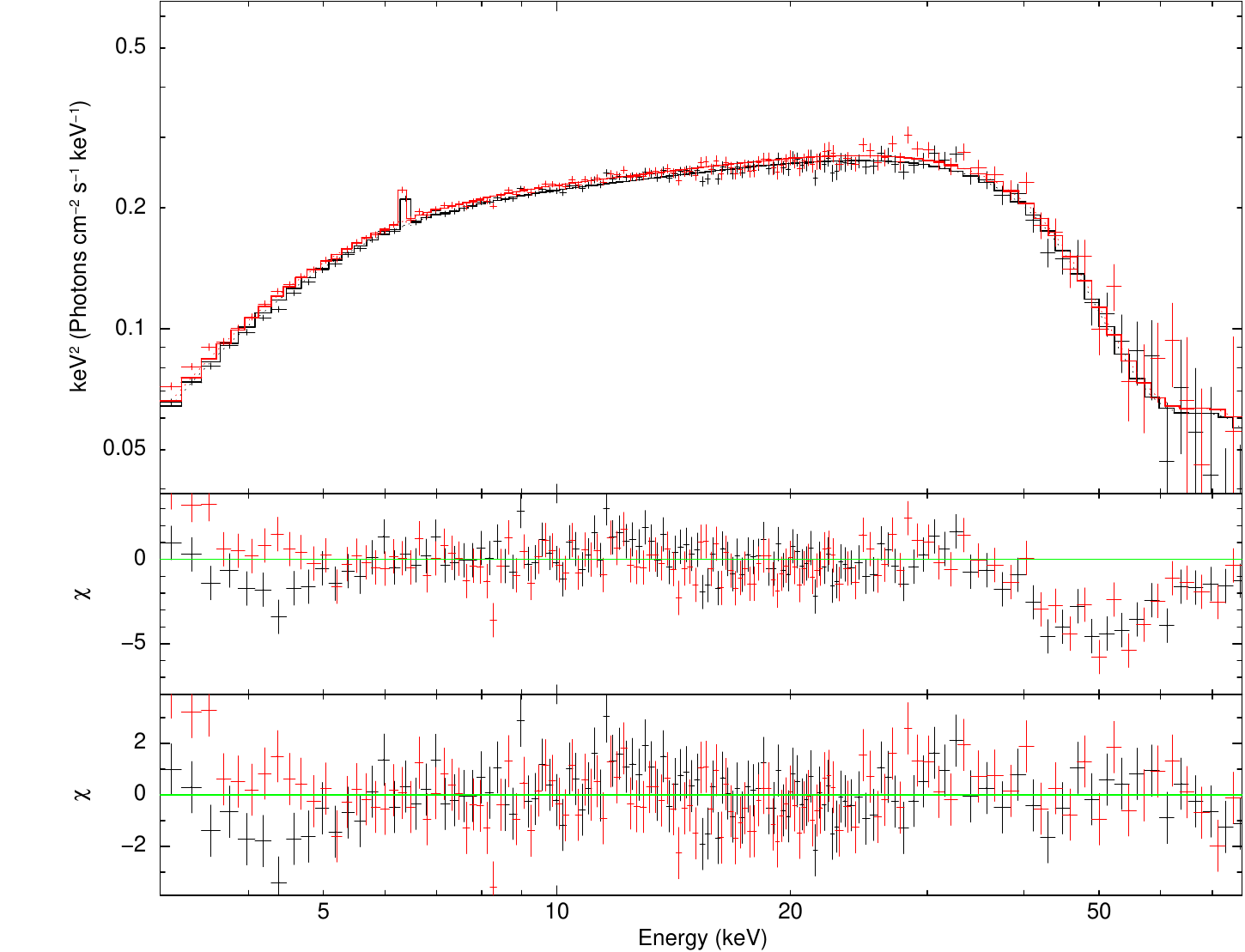}
    \caption{Spectrum and best-fit model for \src\ during NuS-1 ({\it left})) and NuS-2 ({\it right})). Top panels show the best-fit model along with the different components; middle panels show the fit residuals obtained by setting the CRSF strength to zero; bottom panels show the best-fit model residuals. Data points are from \swift/XRT (green), FPMA (black), and FPMB (red).}
    \label{fig:spectrum Nus}
\end{figure*}

The broadband spectrum of \src\ during NuS-1 was modeled using the thermal Comptonization model \texttt{compTT}, which describes the Comptonization of soft photons in a hot plasma ~\citep{comptt_Titarchuk_1994}. The prominent fluorescent iron K$\alpha$ line at 6.4 keV was modeled using a \texttt{Gaussian} component. Initial fits revealed additional spectral residuals near $\sim$6.9 keV and above $\sim$50 keV. The feature near 6.9 keV was modeled with a second \texttt{Gaussian}, however, its width could not be constrained and was therefore fixed to the instrumental resolution. The inclusion of this component gives an improved statistical fit ($\Delta \chi^2 = 98$ for an additional two degrees of freedom). The high-energy residuals above 50 keV are consistent with a CRSF, previously reported in this source \citep{1A1118_Doroshenko_RXTE_2010}. We modeled this feature using the multiplicative absorption component \texttt{gabs}. The CRSF is detected at $56.4^{+1.0}_{-1.0}$ keV with an optical depth of $0.99^{+0.12}_{-0.09}$. 

The final spectral model used was \texttt{tbabs} $\times$ ( \texttt{compTT} + \texttt{Gaussian} + \texttt{Gaussian} ) $\times$ \texttt{gabs}. The best fit values of the spectral parameters are listed in Table~\ref{tab:spectral parameters Nus}. The best-fit spectral model, along with individual components, is shown in the left panel of Fig.~\ref{fig:spectrum Nus}. During NuS-1, the unabsorbed source flux was calculated to be $2.391 \pm 0.004$ $\times 10^{-8}$ ergs$^{-1}$cm$^{-2}$ in the $0.1-100$ keV energy range. Assuming a distance of 2.9 kpc~\citep{GAIA_DR3_2023}, this corresponds to an X-ray luminosity of $2.405\pm0.004$ $\times$ $10^{37}$ erg s$^{-1}$, the brightest luminosity that \src\ has been observed at. The best-fit spectral model yields $\chi^2$/dof=845/744.

A similar spectral model was used for NuS-1. As no simultaneous \swift/\xrt\ data were available during this epoch, the absorption column density could not be constrained and was accordingly frozen to 1.13 $\times$ $10^{22}$ atoms cm$^{-2}$, which corresponds to the Galactic absorption value along the line-of-sight~\citep{Galactic_NH_Kalberla_2005, NH_HI4PI_Collaboration_2016} and similar values were also observed with \swift/\xrt\ during the decay phase of the outburst. The spectrum is well described by the model \texttt{tbabs} $\times$ ( \texttt{compTT} + \texttt{Gaussian} ) $\times$ \texttt{gabs}. The emission line at 6.9 keV is not detected during this observation. The CRSF is clearly detected, with a centroid energy of $58_{-4}^{+6}$ keV and an optical depth of $0.67_{-0.20}^{+0.47}$, and an unabsorbed flux of $1.00 \pm 0.01 \times 10^{-9}$ erg s$^{-1}$ cm$^{-2}$ ($0.1-100$ keV), 
corresponding to X-ray luminosity of $1.01 \pm 0.01$ $\times$ $10^{36}$ erg s$^{-1}$. The best-fit spectral model yielded $\chi^2$/dof=471/439.

\begin{table*}
\centering
\caption{Best-fit spectral parameters of \src\ from joint fitting of \nus\ and \xrt\ data.}
\begin{tabular}{|c|c|c|c|}
\hline
Model & Parameter & Obs-1 + NuS-1 & NuS-2 \\
\hline
tbabs & $N_{\rm H}$ ($10^{22}\rm\,cm^{-2}$) & $3.9_{-0.3}^{+0.3}$& $1.13$ (fixed) \\
\hline
\multirow{4}{*}{compTT} & $T_0$ (keV) & $1.39_{-0.04}^{+0.03}$& $1.29_{-0.01}^{+0.01}$ \\
& $kT$ (keV) & $7.26_{-0.12}^{+0.17}$& $10.8_{-0.7}^{+2.2}$\\
& $\tau_{\rm p}$ & $6.7_{-0.1}^{+0.1}$& $3.15_{-0.34}^{+0.14}$\\
& Norm & $0.202_{-0.003}^{+0.003}$& $0.007_{-0.001}^{+0.001}$ \\ 
\hline
\multirow{2}{*}{Gaussian} & $E_{\rm Fe}$ (keV) & $6.4_{-0.1}^{+0.1}$& $6.4_{-0.1}^{+0.1}$\\
& Norm ($10^{-4}$) & $36.6_{-1.8}^{+1.6}$ & $1.4_{-0.3}^{+0.3}$\\
\hline
\multirow{2}{*}{Gaussian} & $E_{\rm Fe}$ (keV) & $6.9_{-0.2}^{+0.1}$& -- \\
& Norm ($10^{-3}$)& $1.2_{-0.5}^{+0.5}$& -- \\
\hline
\multirow{3}{*}{CRSF} & $E_{\rm cyc}$ (keV) & $56.4_{-1.0}^{+1.0}$& $58_{-4}^{+6}$\\
& $\sigma$ (keV) & $12.5_{-0.9}^{+1.0}$& $9.4_{-2.5}^{+4.2}$\\
& Depth & $0.99_{-0.09}^{+0.12}$& $0.67_{-0.20}^{+0.47}$\\
\hline
Cross-calibration $C_{\rm FPMB}$ &  & $1.01_{-0.01}^{+0.01}$ & $1.03_{-0.01}^{+0.01}$ \\
Cross-calibration $C_{\rm XRT}$ &  & $0.74_{-0.01}^{+0.01}$ & -- \\
\hline
Unabsorbed Flux (0.1–100 keV) & ($10^{-9}$ erg cm$^{-2}$ s$^{-1}$) & $23.91_{-0.04}^{+0.04}$ & $1.00_{-0.01}^{+0.01}$\\
Luminosity (0.1–100 keV) & ($10^{36}$ erg s$^{-1}$) & $24.05_{-0.04}^{+0.04}$ & $1.01_{-0.01}^{+0.01}$\\
$\chi^2$/dof &  & $845/744$& $471/439$\\
\hline
\end{tabular}
\footnote{Errors correspond to the 90\% confidence interval.}
\label{tab:spectral parameters Nus}
\end{table*}

\subsection{Spin-Phase Resolved Spectroscopy} \label{sec: PRS}

We performed spin phase-resolved spectroscopy of the two \nus\ observations to study the variation of the source emission characteristics with the NS spin. For NuS-1, the pulse profile was divided into ten equal phase bins, and the spectra were extracted for each phase segment. As the NuS-2 was 25 times fainter than NuS-1, we divided the data into 4 phase bins. In both cases, the same spectral model used for the phase-averaged analysis was applied to fit each phase-resolved spectrum. 

\begin{figure}
    \centering
    \includegraphics[width=0.48\linewidth]{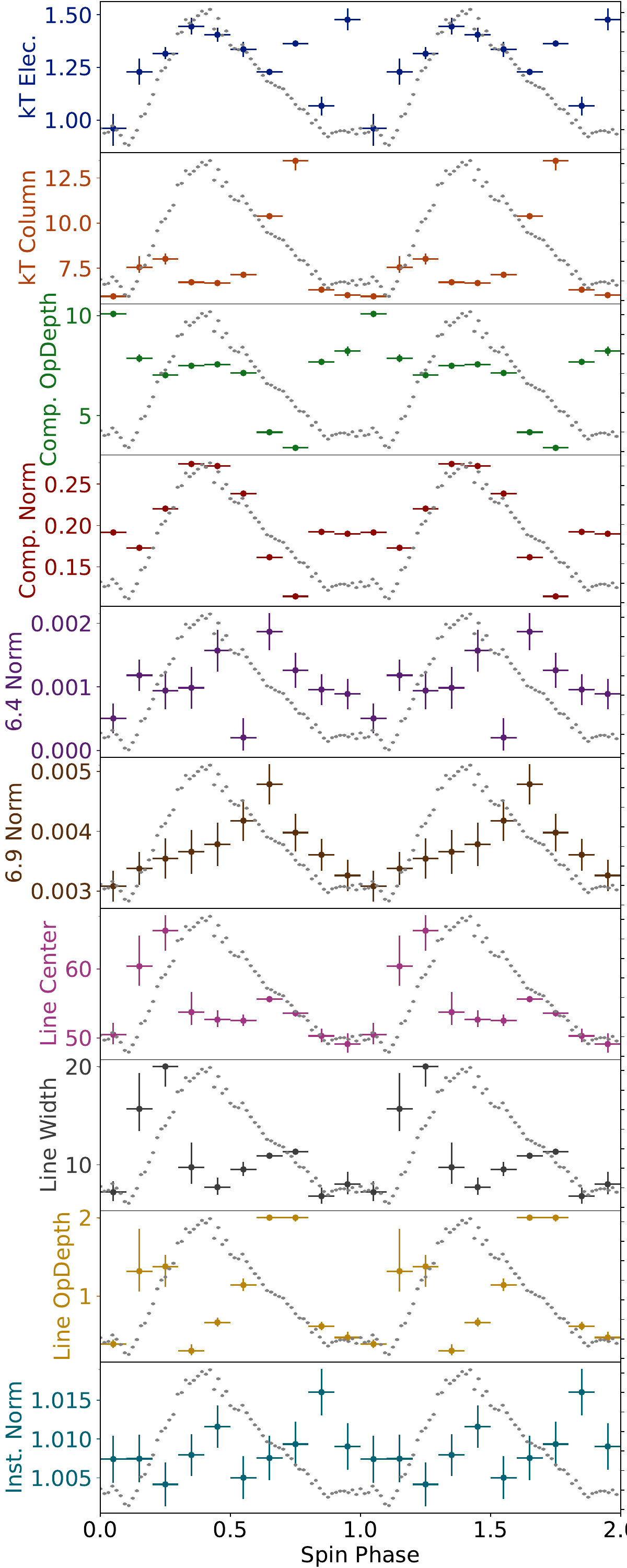}
    \includegraphics[width=0.48\linewidth]{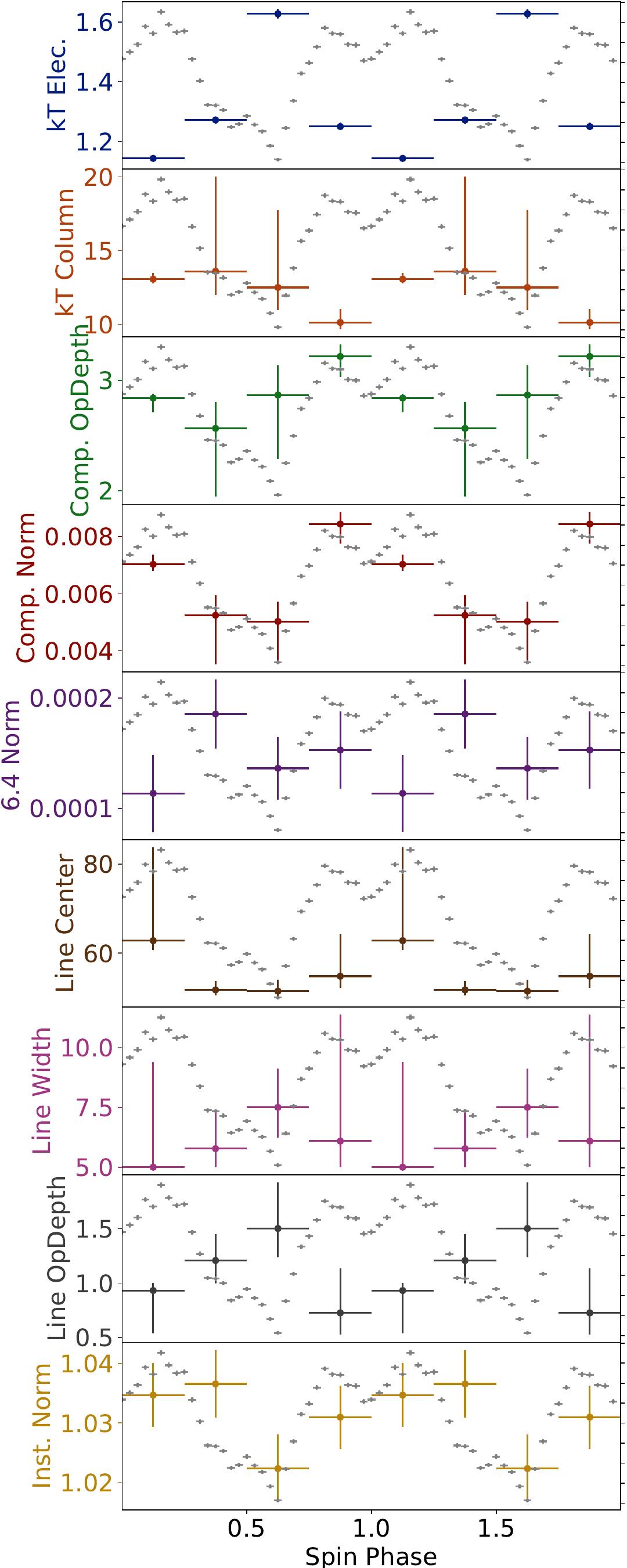}
    \caption{Variations of spectral fit parameters as a function of NS pulse phase, with the pulse profile (grey) in the background, for NuS-1 ({\it Left}) and NuS-2 ({\it right})).}
    \label{fig:nustar_PRS}
\end{figure}

All the spectral parameters show significant variability with the rotation of the NS. Variation of spectral parameters with pulse phase of NS is shown in Fig.~\ref{fig:nustar_PRS}. 
For NuS-1, the CRSF parameters show pronounced pulse phase dependence, with line energy changing by $\sim$15 keV over pulse cycle. The CRSF optical depth shows a similar dependence on pulse phase, as well as a double-peaked profile. The dependence of the CRSF parameters is non-trivial with respect to the NS pulse profile.
During NuS-2, spectral parameters also show significant pulse phase variability (right panel of Fig.~\ref{fig:nustar_PRS}). The CRSF energy exhibits single-peak behavior, varying by around 20 keV. The variation in CRSF line center is significantly different from the NS pulse profile. The optical depth also shows an anti-correlated behavior with the source count rate.

\subsection{Time evolution of spectral parameters} \label{sec: TRS}

Regular monitoring with \swift-\xrt\ allowed us to track spectral evolution during the 2026 outburst. We performed spectral modeling of all \swift-\xrt observations in the $0.7-10$ \rm{keV} band. We initially fitted the spectra with an absorbed power-law model \texttt{tbabs} $\times$ \texttt{po}, which resulted in significant residuals at low energies, indicating the presence of a soft excess. We then introduced an additional blackbody component; however, this led to unphysical parameter values, particularly for the temperature and photon index. Motivated by this, we adopted a partial covering absorption model \texttt{tbabs} $\times$ \texttt{pcfabs}  $\times$ \texttt{po}, which provided statistically acceptable fits for all observations.

\begin{figure*}
    \centering
    \includegraphics[width=0.45\linewidth]{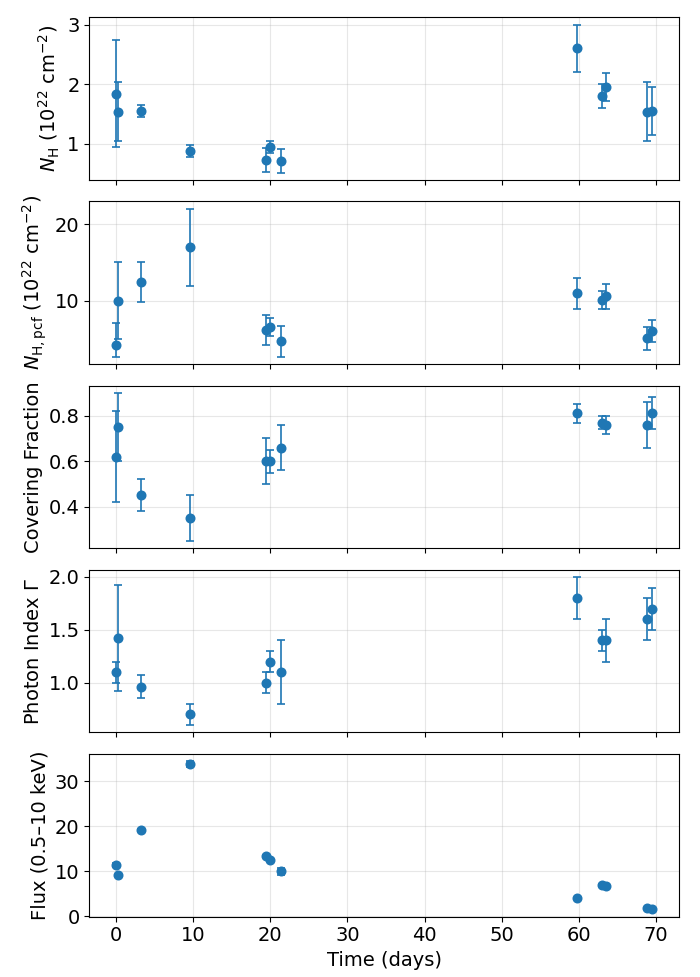}
    \includegraphics[width=0.45\linewidth]{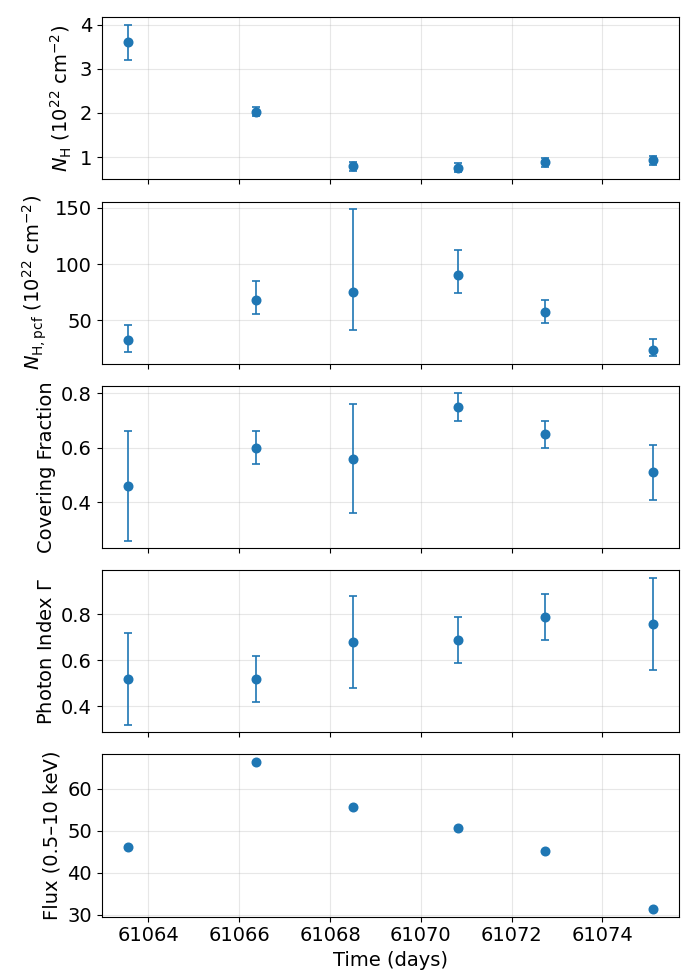}
    \caption{Temporal evolution of spectral parameters during the 2009 (left) and 2026 (right) outbursts as observed with~\xrt. From top to bottom, we show $N_H$, covering fraction, $\Gamma$, and unabsorbed flux in the 0.5--10 keV band, when fitted using a partially covered absorbed power-law model. The results indicate strong variability in both the absorbing material and spectral slope, with the 2026 outburst exhibiting comparatively harder spectra and higher flux levels.}
    \label{fig:spectral-evol-XRT}
\end{figure*}

We also revisited archival \xrt\ observations from the 2009 outburst of \src, previously reported by \citet{Chun-Che2010}. In that work, photon indices as high as $\sim$6 were reported for some observations, which are likely unphysical. We therefore reanalyzed the 2009 \xrt data and performed a uniform analysis of all observations of \src\ from both the 2009 and 2026 outbursts. Fig.~\ref{fig:spectral-evol-XRT} shows the spectral evolution during the 2009 and 2026 outbursts. The top panels indicate variability in the equivalent hydrogen column density ($N_{H}$), suggesting changes in the local absorbing material.

\begin{figure}
    \centering
    \includegraphics[width=\linewidth]{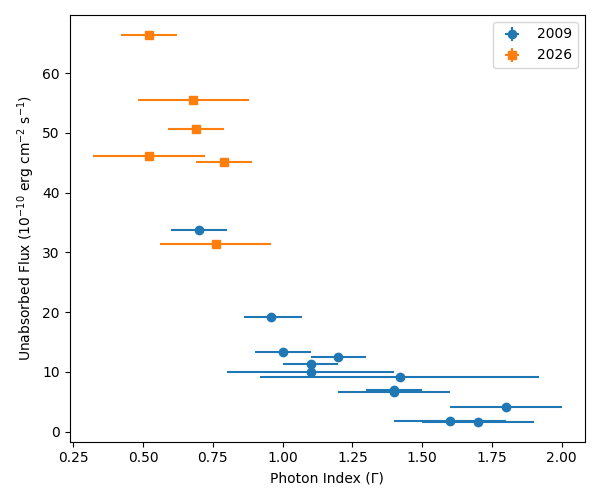}
    \caption{Evolution of the unabsorbed flux (0.5--10 \rm{keV}) as a function of photon index ($\Gamma$) for the 2009 (blue) and 2026 (orange) outbursts observed with~\xrt. The spectral parameters were obtained using a partially covered absorbed power-law model. Error bars represent 1$\sigma$ uncertainties.}
    \label{fig:PL-flux-evol}
\end{figure}

The $N_H$ values did not show any clear correlation with X-ray flux.~The partial covering column density and covering fraction exhibit significant evolution, pointing to a clumpy and dynamic absorber. The photon index varies systematically during both outbursts, indicating spectral state changes.~Notably, the spectra tend to harden (lower $\Gamma$) at higher flux levels.~In Fig.~\ref{fig:PL-flux-evol}, we show the evolution of unabsorbed flux as a function of photon index. A clear distinction is seen between the two outbursts: the 2026 outburst is systematically harder and brighter compared to the 2009 outburst. 

Given that the 2026 outburst exhibits systematically harder spectra, we further explored physical modeling using a thermal Comptonization model \texttt{tbabs} $\times$ \texttt{comptt}.~The best-fit spectral parameters are given in Table~\ref{tab:xrt-comptt}. This model provided an equally good description of the spectra, with electron temperatures ($kT_{\rm{e}}$) varying between 3 and 10~\rm{keV} and high optical depths (13--19).~The seed photon temperature~($kT \sim$ 0.3--0.4~\rm{keV}) remains relatively stable across observations.

\begin{table*}
\caption{Thermal Comptonization Spectral fits to \textit{Swift-XRT} Observations of \src.}
\setlength{\tabcolsep}{3pt}
\begin{tabular}{|c| c| c| c| c| c| c|}
\hline
ObsID & Flux & $N_H$ & kT  & kT$_{\rm{e}}$ & $\tau$ & $\chi^2$/$\nu$   \\
 & [$\times 10^{-10}$ erg cm$^{-2}$ s$^{-1}$] & [10$^{22}$ atom cm$^{-2}$] &  [keV] & [keV]  &   & \\
\hline
00031370006 & $46.2 \pm{0.1}$  & $3.2 \pm^{0.5}_{0.7}$ & $0.4 \pm^{0.3}_{0.4}$ & $3.8 \pm^{0.7}_{0.5}$ & $19 \pm^{4}_{3}$ &   $277/261$ \\
00031370007 & $66.4 \pm 0.1$ & $2.0 \pm{0.1}$ & $0.4 \pm{0.1}$ & $7 \pm{1}$ & $15.3\pm{0.5}$ & $670/640$ \\
00031370008 & $55.6 \pm 0.1$ & $0.7\pm{0.3}$ & $0.4 \pm ^{0.2}_{0.4}$ & $6\pm ^{4}_{1}$ & $13 \pm {1}$  & $192/233$ \\
 00031370009 & $50.6 \pm 0.1$ & $0.7 \pm{0.1}$ & $0.4 \pm{0.1}$ & $10 \pm^{4}_{2}$ & $16 \pm ^{3}_{1}$ & $640/626$ \\
 01442056001 & $45.2 \pm 0.1$ & $0.8 \pm{0.1}$ & $0.35 \pm{0.06}$ & $7 \pm^{2}_{1}$ & $14.0 \pm^{0.9}_{0.5}$ & $625/585$ \\
01442056002 & $31.4 \pm 0.1$ & $0.8 \pm{0.2}$ & $0.3 \pm^{0.1}_{0.3}$ & $3.2\pm{0.5}$ & $16 \pm{2}$ & $244/271$ \\
\hline
\end{tabular}
\footnote{spectral energy range: 0.7-10 keV model used: \texttt{tbabs $\times$ comptt}}
\label{tab:xrt-comptt}
\end{table*}

\section{Discussion} \label{sec: disc}

In this paper, we report on the spectro-temporal properties of the transient BeXRB \src\ during its outburst in early 2026. We utilized simultaneous observations from \nus\ and \xrt\ to perform a broadband spectral and timing analysis of the source. In addition, we used the monitoring capabilities of \xrt\ to investigate the spectral evolution of \src\ during the 2026 outburst and compared it with archival data from the 2009 outburst.

Coherent pulsations were detected with spin periods of $408.77(3)$ \rm{s} and $408.16(1)$ \rm{s} during the two \nus\ observations, indicating a possible spin-up of the NS. The pulse profile evolves significantly during the outburst. We also detect a QPO at $110\pm10$~\rm{mHz} during NuS-1, while no QPOs were detected in NuS-2, suggesting a transient nature. 

The spectrum of \src\ could be well explained by a Comptonized spectrum for both the \nus\ and \swift/\xrt\ data. CRSF were detected in the source from both \nus\ observations, with the line energy changing from $\sim 56$ to $\sim 58$ keV and a change in $L_X$ from 24 to 1 $\times$ $10^{36}$ erg s$^{-1}$. A slight negative correlation between cyclotron line energy and luminosity was therefore seen during the 2026 outburst of \src. We also studied the spectral evolution of \src\ during its 2009 and 2026 outbursts with \swift/\xrt\ (Fig.~\ref{fig:spectral-evol-XRT}), indicating changes in the emission mechanism of the source between the two outbursts.

\subsection{Pulse Profile Evolution}

The \nus\ profiles of~\src\ show a pronounced evolution with luminosity, reflecting changes in the emission geometry of the accretion column. During the high-luminosity (NuS-1) observation, the 3$-$79~\rm{keV} profile is relatively simpler, whereas at lower luminosities (NuS-2) it develops a more structured, double-peaked morphology. The change in pulse morphology observed between the two \nus\ epochs, corresponding to a factor of $\sim25$ decrease in luminosity, is broadly consistent with a transition in emission geometry.~At higher luminosities, the formation of a radiative shock above the NS surface leads to the development of an optically thick accretion column, where radiation preferentially escapes perpendicular to the magnetic field lines, resulting in a fan-beam dominated emission pattern. This naturally produces broader and more structured pulse profiles. As the luminosity decreases, the column becomes optically thinner, and the emission is expected to be more aligned with the magnetic axis, giving rise to a pencil-beam-dominated pattern and simpler pulse profiles~\citep{Basko76}. 

Further support for this scenario comes from the energy-dependent evolution of the pulse profiles. At lower energies, the profiles exhibit multiple peaks and pronounced sub-structures, while at higher energies they tend to simplify and become more single-peaked, accompanied by an increase in pulsed fraction~\citep[see e.g.][]{Nagase89}. This behavior suggests that higher-energy photons originate from regions closer to the NS surface, where the emission is more strongly beamed and less affected by scattering processes in the accretion column~\citep[see e.g.][]{Becker2007}. The increase in pulsed fraction with energy is also consistent with reduced contribution from unpulsed or scattered emission at higher energies, as seen in several accreting pulsars~\citep{Tsygankov2007, Lutovinov2009, RX_J0440p9p4431_QPO_Rahul_2024, Sharma2026}.

The pulse profile behavior observed in the \xrt~data, on the other hand, suggests that the pulse profile evolution cannot be explained solely in terms of luminosity-driven changes in the beaming pattern. During the 2009 outburst, no clear correlation is observed between the X-ray flux and the pulse morphology, while in the 2026 outburst, the profiles consistently exhibit a double-peaked (bi-horned) structure across a range of luminosities. This indicates that additional factors, such as changes in the accretion flow geometry, local absorption, and viewing angle effects, play a significant role in shaping the observed pulse profiles. In particular, the presence of partial covering absorption and spectral variability during the \swift\ observations suggests that phase-dependent obscuration or scattering in the accretion stream may contribute to the formation of complex pulse structures, independent of the global luminosity state~\citep[see e.g.][]{1A1118_Suzaku_Chandreyee_2012}.

\subsection{RMS-Pulse Fraction Evolution}

Interestingly, the pulsed fraction is relatively high at the lower energies, followed by a decrease at intermediate energies ($\sim$5--10 keV). The enhanced pulsed fraction at low energies suggests that the soft emission includes a significant pulsed component, possibly originating from the NS surface or the base of the accretion column. At intermediate energies ($\sim$5--10 keV), the pulsed fraction decreases, likely due to the presence of additional unpulsed emission components such as thermal disk emission or scattered photons, which dilute the intrinsic pulsations, as also inferred from phase-dependent absorption and soft X-ray complexity in accreting pulsars~\citep[see e.g.][]{1A1118_Suzaku_Chandreyee_2012, 1A1118_Suzaku_Suchy_2011}. Alternatively, scattering of high-energy photons in the NS atmosphere could also reduce the amplitude of the pulsed emission \citep{RX_J0440p9p4431_QPO_Rahul_2024}.

Above 10 keV, the pulsed fraction in NuS-1 shows a gradual increase, while in NuS-2 it remains relatively flat, likely reflecting a transition region where multiple emission components contribute with comparable relative strengths \citep{Tsygankov2007}.
At higher energies ($\gtrsim 25$ keV), a significant increase in pulsed fraction is observed, indicating that the emission is increasingly dominated by the Comptonized component. This component is expected to be more anisotropic and therefore more strongly modulated with the NS rotation. The sharper rise in pulsed fraction at high energies in NuS-2 compared to NuS-1 suggests possible changes in the accretion geometry or the relative contribution of the Comptonizing region between the two observations. This is also evident from changes in spectral parameters (Table~\ref{tab:spectral parameters Nus}).

The pulsed fraction during both \nus\ observations shows a prominent increase, reaching a peak around $\sim$50~keV. Interestingly, this energy coincides with the CRSF detected in the X-ray spectrum. Such behavior may indicate a connection between the cyclotron scattering process and the energy-dependent beaming pattern, as variations in scattering cross-section near the cyclotron energy can significantly affect the observed pulsation amplitude~\citep[see e.g.][]{Schonherr2014, Ferrigno2023}.

\subsection{Low-frequency QPOs in \src}

The mHz QPOs in high-mass X-ray binaries (HMXBs) are relatively rare and often transient in nature, and are generally associated with phenomena related to the inner accretion disc \citep[e.g.,][]{QPO_review_Finger_1998, QPO_pulsars_Hemanth_2024, Sharma2025}. 
Several models have been proposed to explain these low-frequency QPOs, including the Keplerian frequency model \citep[KFM;][]{vanderKlis1987}, beat frequency model \citep[BFM;][]{BFM_QPO_Alpar_1985}, 
stochastic viscosity fluctuations in the inner accretion disc \citep{Viscous_flow_QPO_Mushtukov_2019}, and magnetically driven precession of a warped accretion disc \citep{Warped_disc_QPO_Shirakawa_2002}.

According to KFM, accreting inhomogeneity in the accretion disc rotating at the magnetospheric radius with the Keplerian frequency ($\nu_{k}$) produces the QPO ($\nu_{QPO} = \nu_k$). Whereas in BFM, the QPOs result from a beat formation between the orbital frequency of the inner accretion disc ($\nu_{k}$) and the NS spin frequency ($\nu_{NS}$), giving $\nu_{QPO} = \nu_k - \nu_{NS}$. Since the QPO frequency is much higher than the spin frequency, $\nu_{QPO} \approx \nu_k$. The KFM and BFM are indistinguishable and have similar radii at which the QPOs are generated.

The Keplerian radius corresponding to a given frequency is given by
\begin{equation}
    r_{k} = \Big( \frac{GM_{NS}}{4\pi^2 \nu_{k}^2} \Big)^{1/3}, 
    \label{eq:KFM}
\end{equation}
where $M_{NS}$ is the mass of the NS. 

The inner accretion disc is expected to terminate at the magnetospheric radius ($R_m$), where the energy density of the magnetic field
balances the kinetic energy density of the infalling material \citep{Davies1981}. This radius ($R_m$) can be expressed in terms of the luminosity ($L$) and the magnetic moment ($\mu$) of NS as~\citep{Frank_1992},
\begin{equation}
    R_{m} = 3 \times 10^{8} L_{37}^{-2/7} \mu_{30}^{4/7} \quad cm
    \label{eq:magnetospheric radius}
\end{equation}
where $L_{37}$ is in units of $10^{37}$  erg/s and $\mu_{30}$ is in units of $10^{30}$ G cm$^3$. 

Assuming $r_k=R_m$, the observed QPO frequency of 0.11 Hz implies a magnetic field strength of $\sim 7.4 \times 10^{12}$ G, which is within 10\% of the value obtained from the \nus\ cyclotron line. This consistency suggests that the QPOs likely originate near the inner edge of the accretion disc, indicating that the KFM/BFM provides a natural explanation for the observed QPO, favoring an origin associated with inhomogeneities or structures orbiting at the inner disc boundary.

Stochastic viscosity fluctuations in the accretion disc can lead to local variability in the mass accretion rate, which can produce QPOs \citep{Viscous_flow_QPO_Mushtukov_2019}.
The viscous time scale of the accretion flow at a distance (r) from the center of the NS is given by (eq. 5.69 of \cite{Frank_1992}),
\begin{equation}
    t_{visc} (r) = 3 \times 10^5 \alpha ^{-4/5} \dot M_{16} ^{-3/10} M_1 ^{1/4}r_{10}^{5/4} \quad s 
\end{equation}
where $\alpha$ is the viscosity coefficient, $\dot M_{16}$ is the mass accretion rate in units of $10^{16}$ gm \psec, $M_1$ is the mass of the compact object in solar masses, and $r_{10}$ is the distance from the NS centre in units of $10^{10}$ cm. Assuming luminosity is given by $L \sim GM_{NS}\dot M / R_{NS}$ where $M_{NS}$ is the mass of the NS, $R_{NS}$ is the radius of the NS, and $\dot M$ is the mass accretion rate. Using the estimated mass accretion rate of $1.3 \times 10^{17}$ gm \psec\ and assuming the inner disc radius to be close to the magnetospheric radius $7 \times 10^{8}$ \rm{cm}, we obtain a viscous timescale of the order of $10^{4}-10^{5}\,\rm{s}$ for reasonable values of the viscosity parameter (0.001–0.005). This is significantly longer than the observed QPO timescale (9s), indicating that viscous processes are unlikely to directly account for the observed QPO frequency.

The precession of a warped accretion disc due to misaligned disc angular momentum and magnetic field can also cause the presence of QPO signatures~\citep{Warped_disc_QPO_Shirakawa_2002}. The corresponding precession timescale ($\tau_{prec}$) is given by,
\begin{equation}
    \tau_{prec} \sim 776 \times L_{37}^{-0.71} \alpha_{0.1}^{0.85} \quad s
    \label{eq:precession freq}
\end{equation}
$\alpha_{0.1}$ is the viscosity parameter divided by 0.1. For the observed QPOs to be produced by this mechanism, $\alpha$ should have a value of $\sim$0.0011. Although this precession model has been successfully used to describe the mHz QPOs in 4U 0115$+$63 ~\citep{4U0115p63_QPO_JRoy_2019}, Cen X-3 ~\citep{QPO_CenX3_Bachhar_2022}, and SXP 31.0 (Roy et al. under review). The inferred value of the viscosity parameter, $\alpha$, for \src\ is very low, one to two orders of magnitude lower than in sources where this model has successfully explained the QPO frequency. Therefore, the observed $0.11$ Hz QPOs in \src\ are unlikely to be produced by the precession of a wrapped accretion disc.

\subsection{QPO evolution with luminosity}

\begin{figure}
    \centering
    \includegraphics[width=0.9\linewidth]{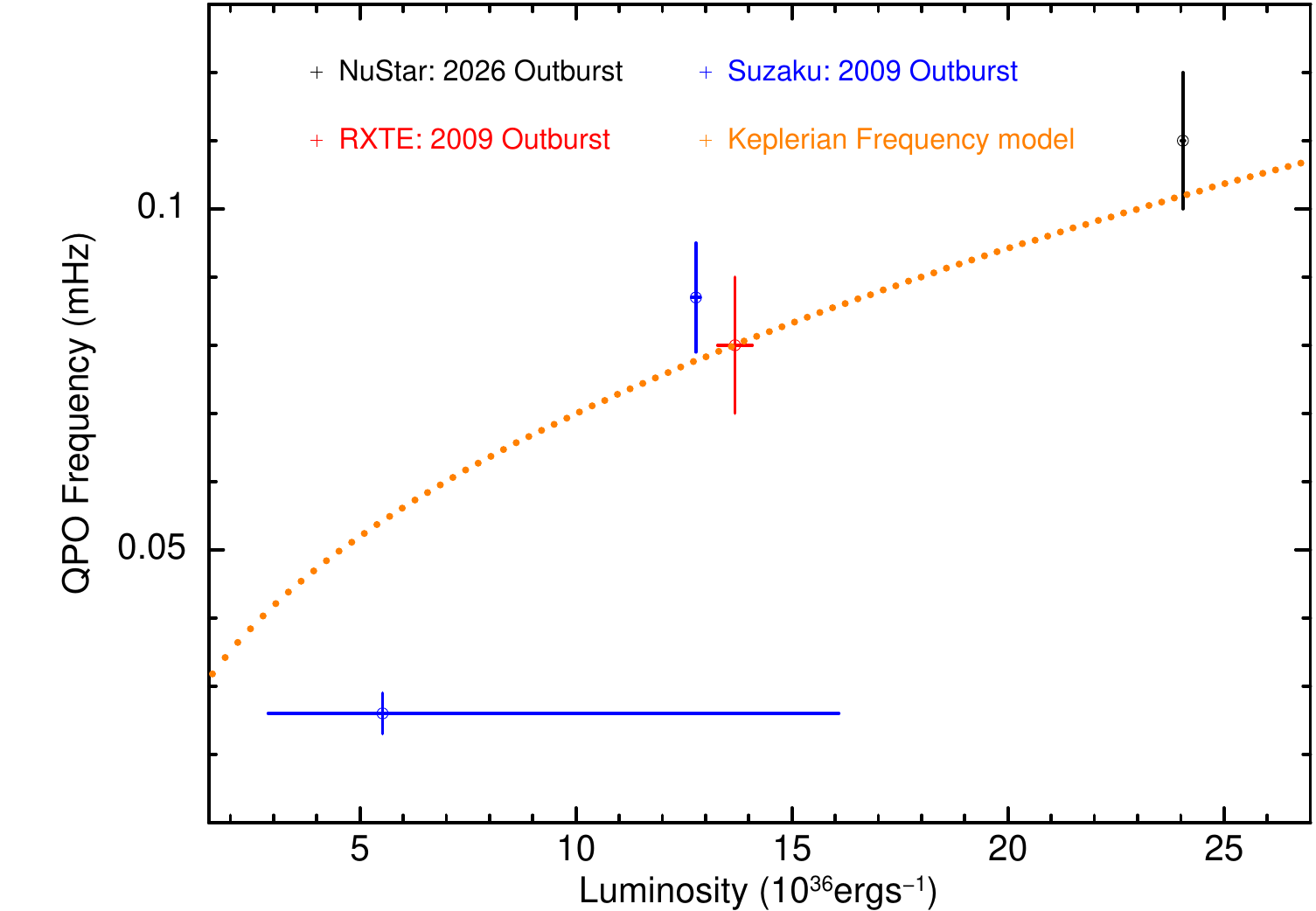}
    \caption{The variation of QPO frequency with luminosity for \src\ is plotted in the figure. The orange points show the predictions from the KFM.}
    \label{fig:QPO vs Luminosity}
\end{figure}

The QPOs in \src\ are known to evolve with source luminosity. The QPO frequency of \src\ during the 2009 outburst evolved from $0.07-0.09$ Hz as seen by RXTE/PCA ~\citep{1A1118_RXTE_PCA_Jincy_2011} and from $0.87-0.026$ Hz as seen with Suzaku~\citep{1A1118_Suzaku_Chandreyee_2012}.  We use the luminosity tagged QPO data available from \rxte\ and \suzaku\ during the 2019 outburst~\citep{1A1118_RXTE_PCA_Jincy_2011, 1A1118_Suzaku_Chandreyee_2012} along with the current QPOs observed in the first \nus\ observation to study its evolution with luminosity (Fig.~\ref{fig:QPO vs Luminosity}). We use the KFM and BFM models to study the dependence of QPO on luminosity. 

We model the QPO frequency dependence on luminosity by combining eq.~\ref{eq:KFM} and eq.~\ref{eq:magnetospheric radius}. We assume a canonical NS of mass of 1.4$M_{\odot}$ and radius of 10 km, and leave the magnetic field strength as a free parameter. The best fit model gives a magnetic field of $6.6\pm0.9$ $\times$ $10^{12}$ G, in good agreement with the value inferred from the CRSF. The corresponding KFM prediction for the QPO frequency as a function of luminosity is shown in orange in Fig.~\ref{fig:QPO vs Luminosity}.

Similarly, we also model the QPO frequency as a function of luminosity using the BFM. This fit gives $B=6.4\pm0.8$ $\times$ $10^{12}$ G, consistent with both the KFM result and the CRSF estimate.
The model successfully reproduces the QPO properties at higher luminosities within a few sigma of the observed data; however, it significantly deviates from the observed QPOs at lower luminosities.
These results indicate that the QPOs most likely originate from instabilities in the accretion flow near the magnetospheric radius.

Using the best-fit luminosity–frequency relation, the predicted QPO frequency during NuS-2 is $\sim$26 mHz, corresponding to roughly one-tenth of the NS spin frequency. This makes the QPO frequency hard to distinguish from the spin harmonics in \src's PDS during NuS-2.

\subsection{Broadband spectra during the 2026 outburst}

The broadband \swift\ and \nus\ spectra of \src\ during the 2026 outburst are well modeled with a thermal Comptonization model. Between two \nus\ observation, the source luminosity decreased by a factor of $\sim$25 from NuS-1 to NuS-2. During NuS-1, both neutral Fe K$\alpha$ (6.4 keV) and ionized Fe XXVI ($\sim$6.9 keV) emission lines are detected, indicating the presence of a highly ionized medium. However, during NuS-2, only the neutral iron line is observed, suggesting a change in the ionization state or location of the line-forming region as the source evolves.

The continuum parameters describing the Comptonized emission remain broadly consistent with those reported during the 2009 outburst \citep{1A1118_Doroshenko_RXTE_2010}. The CRSF energies are also consistent with earlier measurements from  \rxte\ ~\citep{1A1118_Doroshenko_RXTE_2010} and \suzaku\ ~\citep{1A1118_Suzaku_Suchy_2011, 1A1118_Suzaku_Chandreyee_2012}.
CRSF provides a direct estimate of the NS magnetic field strength through the quantization of electron energy levels ($E_n$ in keV) under the influence of strong gravity \citep{Meszaros_1992_book},
\begin{equation}
E_n = \frac{n}{1+z}\frac{eB\hbar}{m_e c} \sim 11.6 \times \frac{n}{1+z} \times B_{12} \quad keV
\label{eq:CRSF}
\end{equation}
where $m_e$ is the electron mass, $B_{12}$ is the magnetic field strength in units of $10^{12}$ G, z is the gravitational redshift, and integer $n$ denotes the Landau levels. With $z$ usually taken as 0.3 for an NS, and assuming that the highest record CRSF energy ($\sim 58$ keV) originates close to the NS surface, we estimate a magnetic field strength of $B \approx 6.5 \times 10^{12}$ G.

\subsection{Spectral evolution between the 2009 and 2016 outburst} \label{subsec: sp evolution}

The spectral evolution of the source was first examined using a phenomenological partial covering power-law model (\texttt{tbabs} $\times$ \texttt{pcfabs}  $\times$ \texttt{po}), which provides an acceptable description of the continuum shape along with variable absorption (Fig.~\ref{fig:spectral-evol-XRT}). The fits reveal significant changes in both $N_H$ and covering fraction.~The observed variability in column density may arise due to changes in the line-of-sight geometry or the presence of clumpy material in the system.~The variation in covering fraction indicates a non-uniform absorber, likely composed of clumpy or structured material. This is consistent with partial covering scenarios often invoked in accreting X-ray pulsars, where the central source is intermittently obscured by local absorbing structures~\citep{Devasia2011, Roy_GX301m2_2024}. A clear anti-correlation between photon index and flux is observed, with the source exhibiting a harder spectrum at higher luminosities (Fig.~\ref{fig:PL-flux-evol}). Such behavior is indicative of enhanced Comptonization, where an increase in electron temperature or optical depth can lead to spectral hardening.

Given that the 2026 outburst was brighter and harder, we further modeled the spectra using a thermal Comptonization model~(Table~\ref{tab:xrt-comptt}). This allows us to probe the properties of the accretion flow, particularly the temperature and optical depth of the Comptonizing medium, and to investigate the origin of the observed spectral evolution. The results from the Comptonization modeling indicate a relatively stable seed photon temperature ($kT \sim 0.3$--$0.4$ keV), suggesting that the origin of soft photons, likely the NS surface or inner accretion disk, does not vary significantly across observations. The electron temperature does not show a clear systematic trend, with values broadly consistent within uncertainties. In contrast, the optical depth exhibits moderate variability, indicating changes in the properties of the Comptonizing medium. The inferred optical depth ($\tau \sim 13-19$) indicates an optically thick Comptonizing medium.

\subsection{Spectral variability with pulse phase}

Phase-resolved spectroscopy of \src\ using \suzaku\ during the 2009 outburst revealed significant variability in spectral parameters over the pulse cycle \citep{1A1118_Suzaku_Chandreyee_2012}. In particular, CRSF showed a variation of $\sim$10 keV in centroid energy and a factor of 3 change in depth. 
In our analysis, the CRSF energy varied by $\sim$15 keV and optical depth by a factor of 5, during NuS-1. 
While CRSF energy changed by about 10 keV, similar to the deviation seen during the \suzaku\ observation, and the optical depth varied by a factor of 2. 
These variations could be attributed to changes in the viewing angle of the cyclotron line-forming region or a more complex underlying magnetic field structure \citep{1A1118_Suzaku_Chandreyee_2012}.

In addition to geometric effects, variations in the CRSF energy may also be influenced by luminosity-dependent changes in the height of the line-forming region within the accretion column. In the super-critical regime, increased radiation pressure can elevate the scattering region to higher elevations, where the magnetic field is weaker, leading to a decrease in the observed CRSF energy. Conversely, in the sub-critical regime, the line-forming region is expected to lie closer to the NS surface, resulting in higher CRSF energies. The observed differences between the two \nus\ observations may therefore reflect changes in the accretion regime in addition to variations in viewing geometry. Similar luminosity-dependent changes in the accretion column structure and emission geometry have been reported in other transient X-ray pulsars such as 4U 1901+03~\citep[see e.g.][]{CRSF_review_Staubert_2019, Beri2020}, supporting the interpretation that variations in the CRSF energy may arise due to shifts in the line-forming region.

\subsection{CRSF evolution with luminosity}

\begin{figure}
    \centering
    \includegraphics[width=0.95\linewidth]{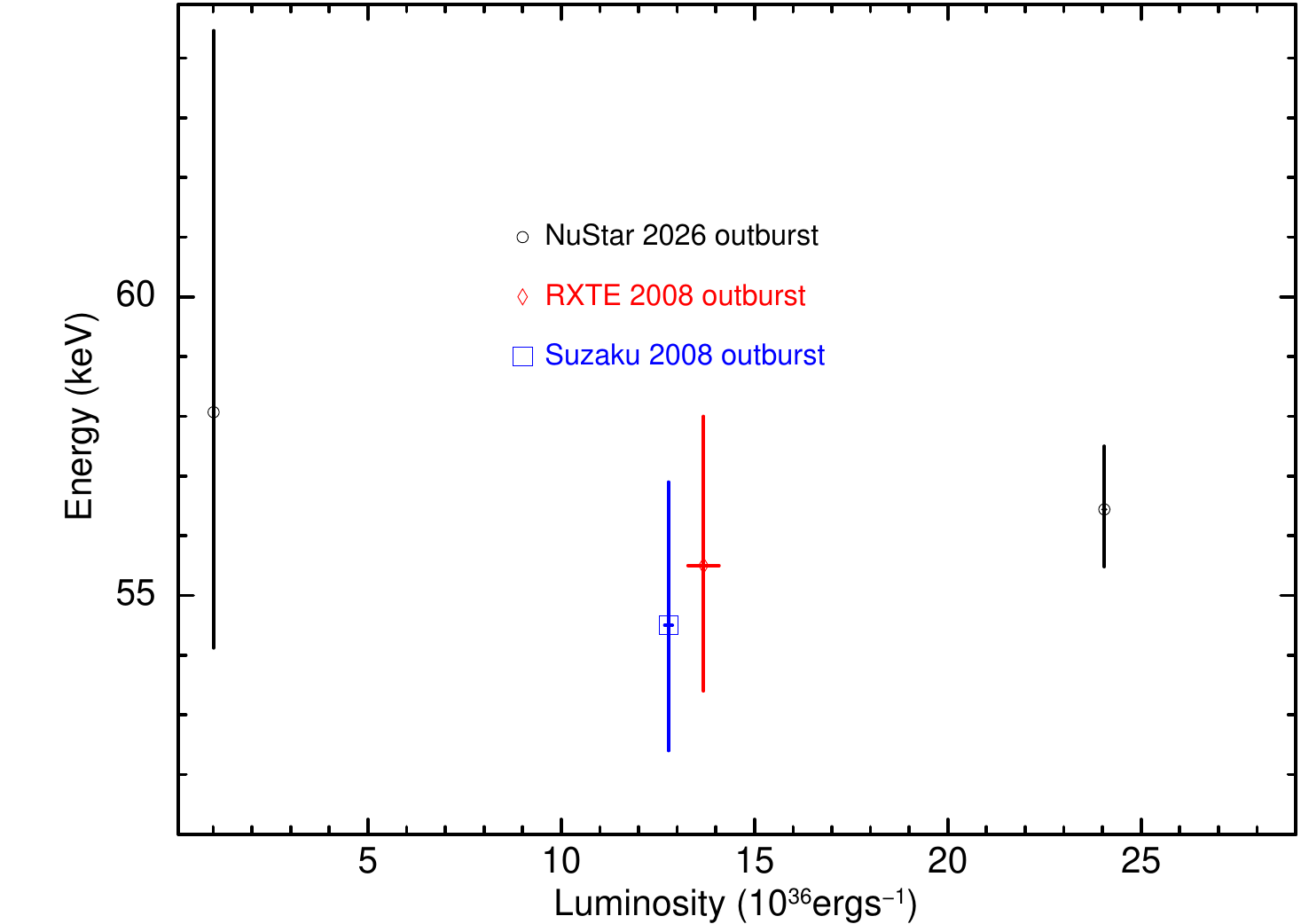}
    \caption{The variation of cyclotron line energy with luminosity for \src.}
    \label{fig:CRSF vs Luminosity}
\end{figure}

To investigate the luminosity dependence of CRSF, we also used previous measurements of CRSF from \rxte~\citep{1A1118_Doroshenko_RXTE_2010} and \suzaku~\citep{1A1118_Suzaku_Suchy_2011} during the 2009 outburst. 
The CRSF centroid energy as a function of luminosity is shown in Fig.~\ref{fig:CRSF vs Luminosity}. There seems to be no correlation between E$_{cyc}$ and L$_X$ across all measurements, covering a luminosity range of $(1-24) \times$ $10^{36}$ erg s$^{-1}$. It is consistent (within errors) with being constant at $\sim$55$-$57keV across the entire luminosity range. This is in contrast with sources such as Cen X$-$3~\citep{CenX3_CRSF_PRS_Tamba_2023} and 4U 0115+63 ~\citep{4U0115_CRSF_PRS_Liu_2020}, where a positive correlation between CRSF and luminosity has been observed. A similar weak or absent correlation has been seen in Cep X$-$4 ~\citep{CRSF_AstroSat_KallolM_2021}, although more recent observations have negated that behavior~\citep{CepX4_CRSF_Roy_2025}. 

The CRSF energy during the 2009 outburst does not show a clear dependence on luminosity within uncertainties, remaining clustered around $\sim$54--56 keV. In contrast, the 2026 observations suggest a possible decrease of CRSF energy with increasing luminosity, although the limited number of measurements prevents a firm conclusion. The difference in behavior between the two outbursts may indicate changes in the location or structure of the line-forming region. This interpretation is further supported by the evolution of the spectral shape across the two outbursts, as revealed by the regular monitoring with \swift\ (Sec.~\ref{subsec: sp evolution}).

The critical luminosity separating sub-critical and super-critical accretion regimes can be estimated using the relation~\citep[see their eq. 32][]{Shock_height_model_Becker_2012}, assuming canonical NS parameters and disk accretion geometry:
\begin{equation}
L_{\mathrm{crit}} \approx 1.5 \times 10^{37} 
\left( \frac{B}{10^{12}\,\mathrm{G}} \right)^{16/15} 
\ \mathrm{erg\ s^{-1}}.
\end{equation}
For a magnetic field strength of $B \sim 6.5 \times 10^{12}$ G, inferred from the CRSF energy, this yields $L_{\mathrm{crit}} \sim (3-5) \times 10^{37}$ erg s$^{-1}$. The two \nus\ observations span nearly an order of magnitude in luminosity, with $L \sim 2.4 \times 10^{37}$ erg s$^{-1}$ for NuS-1 and $L \sim 1 \times 10^{36}$ erg s$^{-1}$ for NuS-2. This places NuS-1 close to the critical regime, while NuS-2 lies firmly in the sub-critical regime. This suggests that the source is observed close to the critical regime during NuS-1, while NuS-2 probes the sub-critical regime. Therefore, a change in the accretion geometry may be the cause of variation in CRSF.

\section{Summary}

In this work, we present temporal and spectroscopic results from two \nus\ observations of \src\ during the 2026 outburst, as well as multiple \swift\ observations during the 2009 and 2026 outbursts, to study variability in its emission properties. The results from the spectroscopic and timing analysis can be summarized as follows :

\begin{itemize}

\item Coherent pulsations at $\sim$400 s are detected across nearly all \nus\ and \swift\ observations.

\item The pulse profiles show strong energy and luminosity dependence, with clear structural changes across the CRSF energy range. The pulsed fraction exhibits complex energy evolution, including a decline beyond the cyclotron line.

\item A cyclotron line is detected at $\sim$56–58 keV in both \nus\ observations, implying a magnetic field strength of $\sim 6.5 \times 10^{12}$ G. Significant pulse-phase variability is observed, with the line energy varying by $\sim$10–15 keV and the optical depth by factors of $\sim$2–5.

\item A QPO at $0.11 \pm 0.01$ Hz is detected during the first \nus\ observation. Its frequency is largely energy-independent, while the rms shows an “elbow-like” energy dependence.

\item The QPO is transient, and its properties and luminosity dependence are consistent with both the KFM and BFM, yielding a magnetic field strength of $\sim 6$–$7 \times 10^{12}$ G, in agreement with the CRSF estimate.

\item The 2026 outburst is significantly harder and more luminous than the 2009 outburst, indicating a stronger contribution from the Comptonizing component.

\item While the CRSF energy does not show strong evolution within individual outbursts, differences between the two outbursts suggest possible changes in the line-forming region and accretion geometry.

\end{itemize}

\begin{acknowledgments}
We thank the Swift and NuSTAR team for accepting our ToO requests and for the quick execution of the observing program. 
KR would like to thank Biswajit Paul for his inputs during the preparation of the manuscript.
\end{acknowledgments}

\facilities{ NuSTAR(FPM), Swift(BAT and XRT), MAXI(GSM)}

\software{astropy \citep{Astropy_2013}, HEASoft, XSPEC \citep{XSPEC}
}

\bibliography{1a1118}{}
\bibliographystyle{aasjournalv7}

\end{document}